\documentclass[a4paper,11pt]{article}
\pdfoutput=1

\usepackage{amssymb,amsmath,bm}
\usepackage{a4wide}
\usepackage{color}
\usepackage{slashed}
\usepackage{graphicx}
\usepackage{amsfonts}
\usepackage{lscape}
\usepackage{hyperref}
\usepackage{amsthm}
\hypersetup{
    colorlinks=true, %set true if you want colored links
    linktoc=all,     %set to all if you want both sections and subsections linked
    linkcolor=blue,  %choose some color if you want links to stand out
    citecolor= blue
}
\usepackage{booktabs}
\usepackage{array}
\usepackage{rotating}
\usepackage[numbers, sort&compress]{natbib}
\usepackage{float}
\usepackage[utf8]{inputenc}
\usepackage[T1]{fontenc}
\usepackage{bm}% bold math
\newcommand{\ord}[1]{\mathcal{O}\left( #1 \right)}

\newcommand{\be}{\begin{equation}}
\newcommand{\ee}{\end{equation}}

\newcommand{\partialLR}{\overset{\leftrightarrow}{\partial^\mu}}
\newcommand{\beq}{\begin{equation}} 
\newcommand{\eeq}{\end{equation}} 
\newcommand{\ba}{\begin{array}}  
\newcommand{\ea}{\end{array}} 
\newcommand{\bea}{\begin{eqnarray}}  
\newcommand{\eea}{\end{eqnarray} }  
\newcommand{\bal}{\begin{align}}
\newcommand{\eal}{\end{align}}   
\newcommand{\bi}{\begin{itemize}}  
\newcommand{\ei}{\end{itemize}}  
\newcommand{\ben}{\begin{enumerate}}  
\newcommand{\een}{\end{enumerate}}  
\newcommand{\bc}{\begin{center}}
\newcommand{\ec}{\end{center}} 
\newcommand{\bt}{\begin{table}}
\newcommand{\et}{\end{table}}  
\newcommand{\btb}{\begin{tabular}}
\newcommand{\etb}{\end{tabular}}

\newcommand{\fref}[1]{Figure~\ref{#1}} 
\newcommand{\eref}[1]{Eq.~(\ref{#1})}

\newcommand{\aref}[1]{Appendix~\ref{#1}}
\newcommand{\sref}[1]{Section~\ref{#1}}

\renewcommand{\Im}{{\rm Im}}  
\renewcommand{\Re}{{\rm Re}} 
\newcommand{\met}{\slashed{E}_T}

\renewcommand{\baselinestretch}{1.2} 

\begin{document}

\vspace{1cm}

\begin{titlepage}

\vspace*{-1.0truecm}
\begin{flushright}
%CERN-TH-2018-068 \\
 \end{flushright}
\vspace{0.8truecm}

\begin{center}
\renewcommand{\baselinestretch}{1.8}\normalsize
\boldmath
{\LARGE\textbf{
Simple model for large CP violation in charm decays, $B$-physics anomalies, muon $g-2$, and Dark Matter
}}
\unboldmath
\end{center}

\vspace{0.4truecm}

\begin{center}
{\bf Lorenzo  Calibbi$\,^a$, Tianjun Li$\,^{b,\,c}$, Ying Li$\,^{d,\,e}$, and Bin Zhu$\,^{d,\,f}$}

\vspace{0.5truecm}

{\footnotesize

$^a${\sl School of Physics, Nankai University, Tianjin 300071, China \vspace{0.15truecm}}

$^b${\sl CAS Key Laboratory of Theoretical Physics, Institute of Theoretical Physics, \\ Chinese Academy of Sciences, Beijing 100190, China \vspace{0.15truecm}}

$^c${\sl School of Physical Sciences, University of Chinese Academy of Sciences,\\
No.~19A Yuquan Road, Beijing 100049, China  \vspace{0.15truecm}}

$^d${\sl Department of Physics, Yantai University, Yantai 264005, China \vspace{0.15truecm}}

$^e${\sl Center for High Energy Physics, Peking University, Beijing 100871, China \vspace{0.15truecm}}

$^f${\sl Department of Physics, Chung-Ang University, Seoul 06974, Korea}
}

\vspace*{5mm}
%\today
\end{center}

\vspace{0.4cm}
\begin{abstract}
\noindent
We present a minimal extension of the Standard Model that can simultaneously account for the anomalies in semi-leptonic $B$ meson
decays and the muon $g-2$, give large CP violation in charm decays (up to the value recently measured by LHCb),
and provide thermal-relic dark matter, while evading all constraints set by other flavour observables, LHC searches, and dark matter experiments. This is achieved by introducing only four new fields: a vectorlike quark, a vectorlike lepton, and two scalar fields (a singlet and a doublet) that mix due to the electroweak symmetry breaking and provide the dark matter candidate. The singlet-doublet mixing induces chirally-enhanced dipole transitions, which are crucial for the explanation of the muon $g-2$ discrepancy and the large charm CP violation, and allows to achieve the observed dark matter density in wide regions of the parameter space.
\end{abstract}

\end{titlepage}
\tableofcontents
%
%
%\newpage
%\renewcommand{\theequation}{\arabic{section}.\arabic{equation}} 
%
\vspace{-0.2cm}
\section{Introduction}
%
%In this paper, we adopt a bottom-up approach to new physics (NP) beyond the Standard Model (SM) of particle physics. Instead of focusing on UV-complete, theoretically-motivated, NP scenarios (e.g.~addressing the hierarchy problem, grand unification, etc.), we just concern ourselves 
%with a simplified model that can accommodate a number of observational hints for new physics at (or not far above) the TeV scale. 
Instead of focusing on UV-complete, theoretically-motivated, new physics (NP) scenarios (e.g.~addressing the hierarchy problem, grand unification, etc.), we adopt here a bottom-up approach to NP beyond the Standard Model (SM) of particle physics, and just concern ourselves 
with a simplified model that can accommodate a number of observational hints for NP at (or not far above) the TeV scale. 
In fact, although the LHC experiments could not establish the existence of new particles beyond the SM, we have been witnessing in recent years to several persisting discrepancies between observations and SM predictions, especially in the flavour sector.   
One is the muon anomalous magnetic moment, muon $g-2$, which features a long-standing disagreement between theoretical predictions and experiments at the level of more than $3\sigma$. If confirmed, possibly by the results of the new Muon g-2 experiment at Fermilab \cite{Grange:2015fou}, this discrepancy would unambiguously require new particles interacting with muons at the TeV scale or below: cf.~\cite{Lindner:2016bgg} for a review. 
The physics of the $B$ mesons provides other examples. The LHCb and the $B$-factory experiments have observed hints of Lepton Flavour Universality (LFU) violation in semi-leptonic $B$ decays, especially in the observables $R_{K^{(*)}}\equiv  {\rm BR}(B \to K^{(*)} \mu\mu) /{ {\rm BR}(B \to K^{(*)}ee)}$ that are theoretically very clean in the SM (and whose values are predicted to be practically one). In addition to this, semi-leptonic $B$ decay data (again from $b\to s\mu\mu$ processes) exhibit a coherent pattern of observables in tension with the SM, namely a general deficit in the differential branching fractions as well as discrepancies in angular observables. Reviews of such `$B$-physics anomalies' can be found in \cite{Albrecht:2018vsa,Li:2018lxi,Bifani:2018zmi}. 
Besides flavour observables, the evidence of cold dark matter (DM) in the universe could be a further hint for a low-energy NP sector. 
This follows from the `WIMP miracle', the remarkable observation that, assuming a standard thermal history of the universe, the DM relic density measured from observations of the Cosmic Microwave Background (CMB) can be accounted for by particles in the mass range of the electroweak-breaking scale, annihilating with a cross section of the typical electroweak size, cf.~the reviews \cite{Bertone:2004pz,Feng:2010gw}. This motivates the possibility that DM is a so-called Weakly Interacting Massive Particle (WIMP) and can thus be produced and observed (possibly in association with other new particles) at colliders. 

As mentioned above, following a bottom-up approach we want to build and study a minimal model that can simultaneously account for the above hints of new physics. We regard this as a useful exercise to highlight the building blocks that a fully-fledged theory (possibly addressing other major shortcomings of the SM, such as the generation of neutrino masses, the origin of the fermion mass hierarchies, baryogenesis etc.)~may incorporate if the above observations will be proven to be indeed due to beyond the SM dynamics.
We build on previous attempts \cite{Sierra:2015fma,Belanger:2015nma,Megias:2017dzd,Kawamura:2017ecz,Kowalska:2017iqv,Cline:2017qqu,Bian:2017xzg,Calibbi:2018rzv,Li:2018rax,Li:2018aov,Barman:2018jhz,Grinstein:2018fgb,Baek:2019qte,Li:2019xmi} to address (some of) the above experimental results by adding to the SM a limited number of new fields, focusing on heavy scalars and heavy quarks and leptons in vectorlike representations of the SM gauge group (for general discussions of this kind of one-loop solutions of the  $B$-physics anomalies see \cite{Gripaios:2015gra,Arnan:2016cpy,Arnan:2019uhr}). In particular, we extend the model discussed in \cite{Kawamura:2017ecz,Grinstein:2018fgb} by adding a scalar $SU(2)_L$ doublet. The mixing of such field with a scalar singlet (via a coupling with the Higgs) introduces chirally-enhanced dipole transitions that allow to account for the muon $g-2$ with a heavy enough NP spectrum that can be compatible with LHC constraints and the observed DM abundance without the need of tuning the model's parameters, as extensively discussed in \cite{Calibbi:2018rzv}.
This crucial novel ingredient also generates enhanced dipole operators in the quark sector, which can lead to other desirable effects. In particular, we contemplate here the possibility that CP violation in charm decays, which has been recently established by LHCb \cite{Aaij:2019kcg}, is also a NP effect and is accounted for by our simple model.

%In the rest of the paper, after presenting the model in \sref{sec:model}, we thoroughly study its phenomenological implications. In \sref{sec:flavour}, we discuss the flavour effects we are interested in and the relevant constraints set by other flavour observables. In \sref{sec:DM-LHC}, we discuss in detail LHC and DM phenomenology of our model and we combine it with the flavour constraints. We conclude in \sref{sec:conclusions}, while we present some useful formulae in the Appendices.

After presenting the model in \sref{sec:model}, we throughly study its phenomenological implications. In \sref{sec:flavour}, we discuss the flavour effects we are interested in and the relevant constraints set by other flavour observables. In \sref{sec:DM-LHC}, we discuss in detail LHC and DM phenomenology of our model and we combine it with the flavour constraints. We conclude in \sref{sec:conclusions}, while we present some useful formulae in the Appendices.

\section{Field content and interactions}
\label{sec:model}

We introduce the following set of new fields that are all odd under an unbroken $\mathbf Z_2$ symmetry under which the SM fields are even:
a singlet complex scalar, a complex scalar doublet, and two vectorlike pairs of Weyl fermions (that combine into two Dirac fermions) with the quantum numbers of the SM quark and lepton doublets. To summarise, the gauge quantum numbers of the extra fields are as follows:
\begin{center} \begin{tabular}{ccccc}
\hline 
 Field & spin & $SU(3)_c$ & $SU(2)_L$ & $U(1)_Y$ \\
\hline 
$Q^\prime$ & 1/2 & {\bf 3} & {\bf 2} & 1/6 \\
$L^\prime$ & 1/2 & {\bf 1} & {\bf 2} & -1/2 \\ 
$\Phi_S$ & 0 & {\bf 1} & {\bf 1} & 0 \\ 
$\Phi_D$ & 0 & {\bf 1} & {\bf 2} & -1/2 \\ 
 \hline
 \end{tabular}\end{center}
In terms of $SU(2)_L$ components the Dirac fermions can be written as:
\begin{align}
Q^\prime = \begin{pmatrix}
U^\prime \\ D^\prime
\end{pmatrix}, \quad \quad
L^\prime = \begin{pmatrix}
L^{\prime\, 0} \\ L^{\prime -}
\end{pmatrix}.
\end{align}
Given the unbroken $\mathbf Z_2$ that we assumed, these fields do not mix with the SM fermions.
For the same reason, the scalars do not mix with the SM Higgs, although they interact with it via trilinear and quartic `Higgs portal' couplings. The scalar sector can be decomposed as follows:
\begin{align}
\Phi_S \equiv S^0_s\,, \quad \quad 
\Phi_D = \begin{pmatrix}
%(S^0_d + i A^0)/{\sqrt{2}}\\ S^-
S^0_d\\ S^-
\end{pmatrix}, 
\end{align}
The physical states are thus two neutral and one charged complex scalar.

The part of the Lagrangian involving the new fields is given by the following expression:
\begin{align}
\label{eq:lag}
{\cal L}  \supset & \left( \lambda^{Q}_{i}\, \overline{Q^\prime} Q_i \Phi_S + \lambda^{U}_{i}\, \overline{Q^\prime} U_i {\Phi}_D  + \lambda^{D}_{i} \,\overline{Q^\prime} D_i {\widetilde \Phi}_D  + % \right. \nonumber \\ & \left.
 \lambda^{L}_{i}\, \overline{L^\prime} L_i \Phi_S + \lambda^{E}_{i}\, \overline{L^\prime} E_i {\widetilde \Phi}_D  +   a_H \,H^\dagger  \widetilde{\Phi}_D \Phi_S + {\rm h.c.} \right)  \nonumber \\
& -  M_{Q} \overline{Q^\prime} Q^\prime -  M_{L} \overline{L^\prime} L^\prime  - {M_{S}^2} \Phi^*_S \Phi_S - M_{D}^2 \Phi_D^* \Phi_D \, ,
\end{align} 
where we omitted the quartic couplings of the scalar potential, we defined $ {\widetilde \Phi}_D \equiv   i \sigma_2 \Phi_D^*$, 
and we denoted the left-handed (LH) and right-handed (RH) SM fermions respectively as 
$Q_i$,~$L_i$, and ~$U_i$,~$D_i$,~$E_i$, with $i=1,3$ being a flavour index. 
\begin{figure}[t]
\centering
\includegraphics[width=0.8\textwidth]{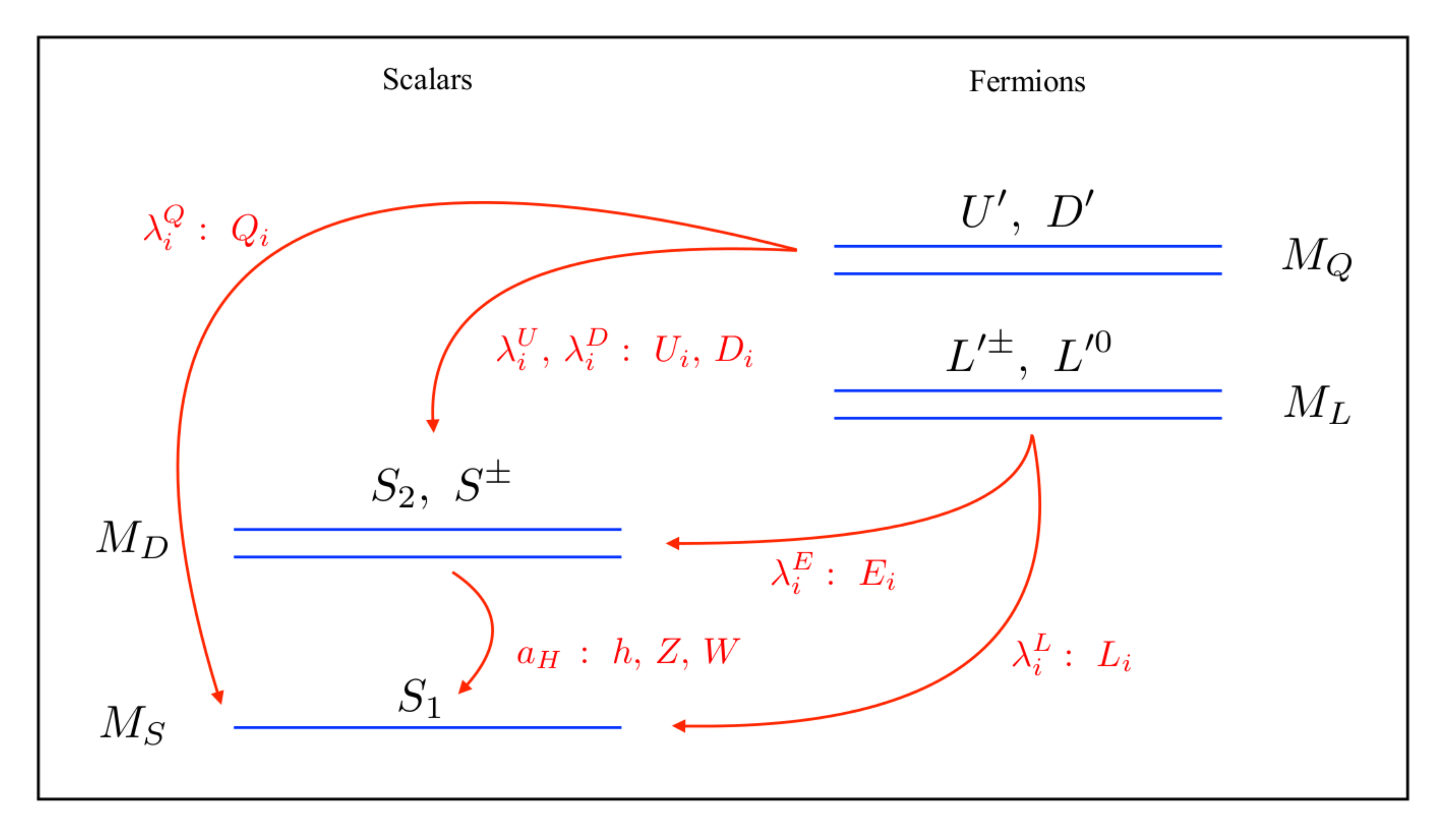}
\caption{\label{fig:spectrum} Schematic representation of the new particles and interactions.}
\end{figure}

Upon electroweak-symmetry breaking, the scalar coupling $a_H$ (that has the dimension of a mass) induces mixing between the neutral components of $\Phi_S$ and $\Phi_D$. The mass matrix and our definition of the mixing matrix $U$ are the following: 
\begin{align}
U^\dagger \begin{pmatrix} M_S^2 &  a^*_H v/\sqrt{2}  \\  a_H  v/\sqrt{2}&  M_D^2 \end{pmatrix} U = \begin{pmatrix} M_{S_1}^2 & \\ & M_{S_2}^2  \end{pmatrix} \, ,
\end{align}
where $v$ is the Higgs field vev $\simeq 246$ GeV. We denote the mass eigenstates as $S_1$ and $S_2$ and by convention we take $M_{S_1}^2 \le M_{S_2}^2 $. Physical masses and mixing are then given by
\begin{align}
M_{S_{1,2}}^2 & = (M_S^2 + M_D^2 \mp \Delta M^2)/2,\quad\quad \Delta M^2 \equiv \sqrt{(M_D^2 - M_S^2)^2 + 2 a_H^2 v^2}  \,,
\end{align}
\begin{align}
\label{eq:Umix}
U  = \begin{pmatrix}
\frac{\sqrt{2} a_H v}{\sqrt{(M_D^2 - M_S^2 - \Delta M^2)^2 + 2 a_H^2 v^2}} &
 -\frac{M_D^2 - M_S^2 - \Delta M^2}{\sqrt{(M_D^2 - M_S^2 - \Delta M^2)^2 + 2 a_H^2 v^2}} \\
\frac{M_D^2 - M_S^2 - \Delta M^2}{\sqrt{(M_D^2 - M_S^2 - \Delta M^2)^2 + 2 a_H^2 v^2}}&
 \frac{\sqrt{2} a_H v}{\sqrt{(M_D^2 - M_S^2 - \Delta M^2)^2 + 2 a_H^2 v^2}}
\end{pmatrix}.
\end{align}
Notice in particular that the entry $U_{1\alpha}$ ($U_{2\alpha}$) represents the singlet (doublet) component in the mass eigenstate $S_\alpha$,
namely: $S_\alpha = U_{1\alpha} S^0_s +  U_{2\alpha} S^0_d$.
If lighter than the vectorlike fermions $S_1$ is a good candidate for cold dark matter, as we will discuss in the \sref{sec:DM-LHC}.\footnote{Quartic interactions in the scalar potential can introduce a mass splitting between the CP-odd and CP-even components of $S_\alpha$, see e.g.~\cite{LopezHonorez:2006gr}. We are going to assume that this is a small effect and ignore it in the discussion of the flavour phenomenology. Such a mass splitting does however play an important role for DM direct detection, cf.~\sref{sec:DM}.} 
Finally, the charged scalar mass is at the tree level simply given by the mass parameter of the scalar doublet:  $M_{S^\pm} = M_D$.\footnote{Electroweak breaking effects induced by singlet-doublet mixing can be in principle constrained by electroweak precision observables. 
However, as we will see in the following sections, the flavour observables we are interested in only require mild values of the mixing parameter $a_H$, hence the impact on electroweak precision observables is negligible. Cf.~the detailed discussion in \aref{sec:ewpo}.}

Another effect of the electroweak breaking is that the term $\lambda^{Q}$ in \eref{eq:lag} induces couplings of the components of the vectorlike quark $Q^\prime$ to LH up and down quarks that we denote respectively as $\lambda^{Q_u}$ and $\lambda^{Q_d}$. These couplings have a relative misalignment in the flavour space given by a CKM rotation. For instance, we can choose a basis such that:
\begin{align}
\label{eq:lambdaQ}
\lambda^{Q_u}_{i} = \lambda^{Q}_{i}\,,\quad\quad \lambda^{Q_d}_{i} = \sum_k\lambda^{Q}_{k}~V^*_{ki}\,,
\end{align}
where $V$ is the CKM matrix.

The Lagrangian written in terms of the mass eigenstates can be found in the Appendix \ref{app:lag}. In \fref{fig:spectrum}, we sketch the spectrum of the new particles and their interactions, assuming for illustration purposes the hierarchy $M_Q > M_L > M_D > M_S$, and a moderate scalar mixing, so that $M_{S_2} \approx M_{S^\pm} = M_D$ and $M_{S_1} \approx M_S$.

\section{Flavour observables and phenomenology}
\label{sec:flavour}
The purpose of this section is to illustrate how the new fields of the model contribute to the flavour observables we are interested in, and to discuss the relevant constraints. This discussion also allows us to identify the interactions (and quantify their strength) that lead to the desired effects. The resulting constraints and benchmark values of the couplings will be employed in the following section in order to study parameter space and spectrum compatible with 
the $B$-physics anomalies, CPV in charm decays, muon $g-2$, and dark matter.

\subsection{LFU violation in $b\to s\ell\ell$ transitions and $B$-physics constraints}
\label{sec:Bphys}
\begin{figure}[t]
\centering
\includegraphics[width=0.48\textwidth]{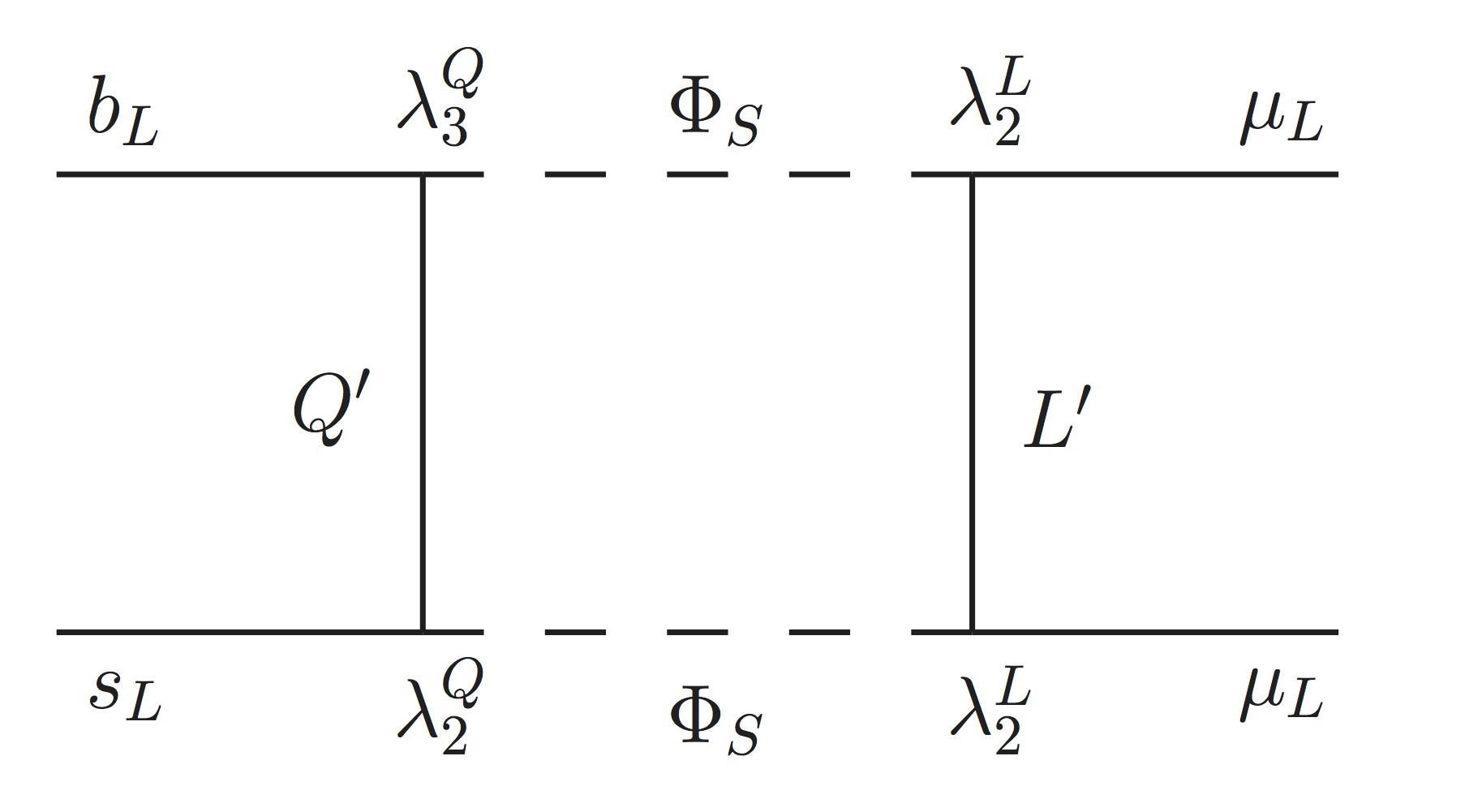}
\includegraphics[width=0.48\textwidth]{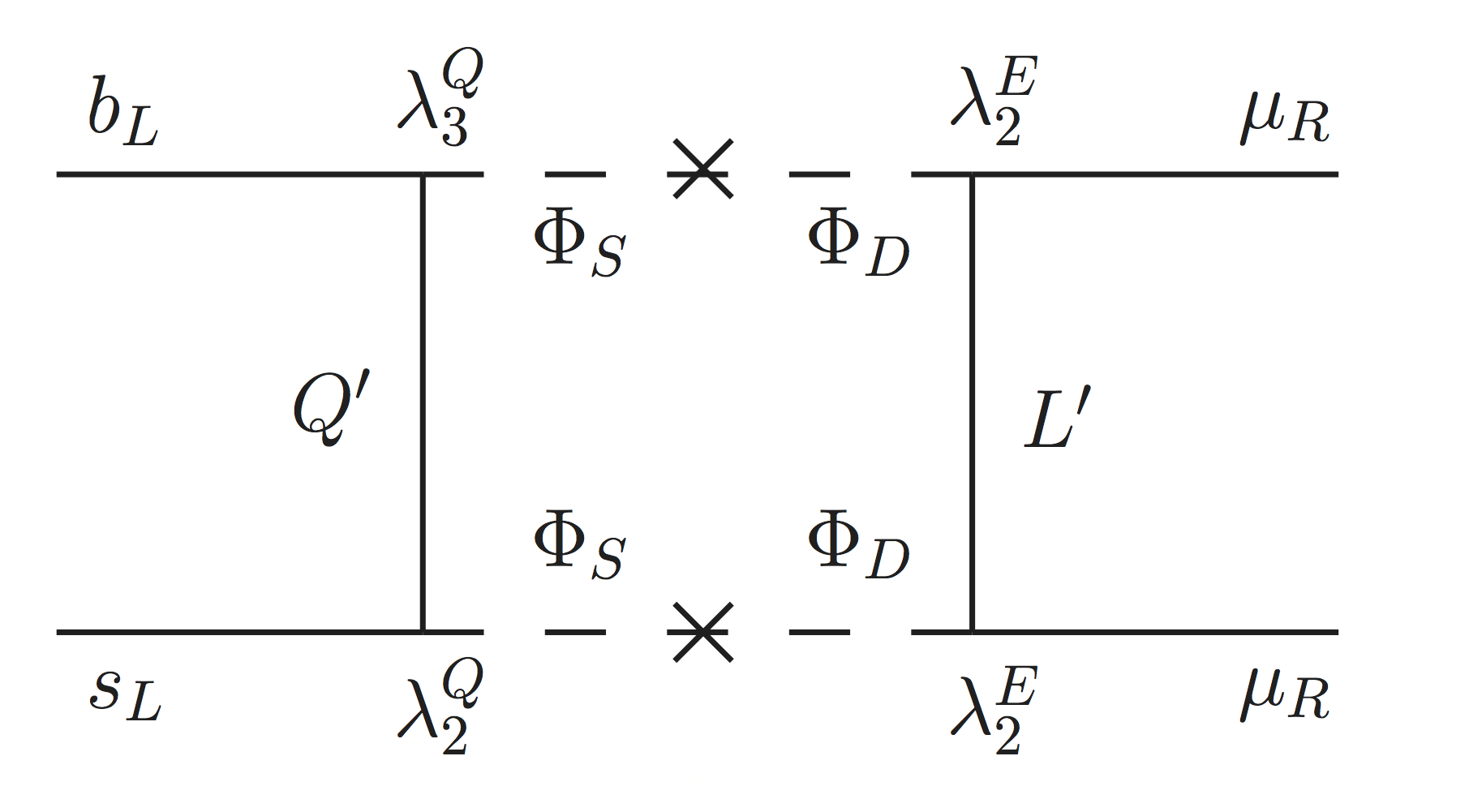}
\caption{Diagrams contributing to the $b\to s \mu \mu$ operators $C_9^{bs\mu\mu}$ and $C_{10}^{bs\mu\mu}$ .\label{fig:bsll}}
\end{figure}

The simplest way to address the anomalies observed in the semi-leptonic $B$ decays LFU observables $R_{K^{(*)}}$ and in branching ratios and angular distributions of several $b\to s\mu\mu$ modes is adding non-standard contributions to the following operators (for the latest fits see \cite{Alguero:2019ptt,Alok:2019ufo,Ciuchini:2019usw,Aebischer:2019mlg,Kowalska:2019ley,Arbey:2019duh,Bhattacharya:2019dot}):
\begin{align}
\mathcal{H}^{bs\mu\mu}_{\rm eff} \supset - \mathcal N \left[C_9^{bs\mu\mu} \,(\overline{s}\gamma_\mu P_L b) (\overline{\mu} \gamma^\mu\mu)+ C_{10}^{bs\mu\mu}\, (\overline{s}\gamma_\mu P_L b) (\overline{\mu} \gamma^\mu\gamma_5 \mu)	  +{\rm h.c.} \right],
\end{align}
where the Wilson coefficient are normalised by the SM contribution
\begin{align}
\mathcal{N} \equiv \frac{4 G_F}{\sqrt{2}} \frac{e^2}{16\pi^2}V_{tb} V_{ts}^*\,.
\end{align}
Adapting to our specific model the formulae of~\cite{Arnan:2019uhr} (see also~\cite{Kawamura:2017ecz}), we get for the contribution to $C_{9,10}^{bs\mu\mu}$ from diagrams involving $Q^\prime$, $L^\prime$, and the scalars $\Phi_S$ and $\Phi_D$ (shown in \fref{fig:bsll}):
\begin{align}
\Delta C_9^{bs\mu\mu}&=-\frac{\lambda_{3}^{Q_d} \lambda_{2}^{Q_d\, *}}{128 \pi^2 \mathcal{N}} \sum_{\alpha=1,2} \frac{\left|U_{1 \alpha}\right|^{4}\left|\lambda_{2}^{L}\right|^{2}+\left|U_{1 \alpha}\right|^{2} \left|U_{2 \alpha}\right|^{2}\left|\lambda_{2}^{E}\right|^{2}}{M_{S_{\alpha}}^{2}} F_2\left(\frac{M_{Q}^{2}}{M_{S_{\alpha}}^{2}}, \frac{M_{L}^{2}}{M_{S_{\alpha}}^{2}}\right),
\\
\Delta C_{10}^{bs\mu\mu}&=\frac{\lambda_{3}^{Q_d} \lambda_{2}^{Q_d\, *}}{128 \pi^2 \mathcal{N}} \sum_{\alpha=1,2} \frac{\left|U_{1 \alpha}\right|^{4}\left|\lambda_{2}^{L}\right|^{2}-\left|U_{1 \alpha}\right|^{2} \left|U_{2 \alpha}\right|^{2}\left|\lambda_{2}^{E}\right|^{2}}{M_{S_{\alpha}}^{2}} F_2\left(\frac{M_{Q}^{2}}{M_{S_{\alpha}}^{2}}, \frac{M_{L}^{2}}{M_{S_{\alpha}}^{2}}\right),
\end{align}
where the loop function is
\begin{align}	
F_2(x,y)\equiv \frac{1}{(x-1)(y-1)} + \frac{x^2 \log x}{(x-1)^2(x-y)}+ \frac{y^2 \log y}{(y-1)^2(y-x)}\,.
\end{align}
Notice that the second terms of both expressions come from the second diagram in \fref{fig:bsll} (involving RH muons) and vanish in absence of scalar singlet-doublet mixing. In such a case the contribution of our model takes the form $\Delta C_9^{bs\mu\mu} = -\Delta C_{10}^{bs\mu\mu}$, typical of new physics coupled to LH leptons only.\footnote{In addition, notice that an equivalent diagram involving one neutral scalar and one charged scalar $S^\pm$ contained in the doublet can give rise to charged-current modes, such as $b\to c\tau\nu$. Nevertheless, these processes arise at the tree level in the SM, hence our loop-suppressed contribution would not be enough to address the anomalies in $R_{D^{(*)}}$, cf.~\cite{Bifani:2018zmi} for a review.}

As it is apparent from the expressions given in the Appendix \ref{app:bsll}, contributions to additional $b\to s\ell\ell$ operators involving RH quarks (and to dipole operators as those in Appendix \ref{app:bsg}) depend on the couplings to RH down quarks, $\lambda^D_i$, and are thus suppressed if such couplings are small. In the following we are going to assume that this is the case (i.e.~$\lambda^D_i \ll \lambda^Q_i$), although some degree of RH currents may help fitting the 
$b\to s \ell \ell$ data (see e.g.~\cite{Alguero:2019ptt}). This choice is also motivated by the constraint from $b\to s\gamma$ transitions, $B_s\to \mu\mu$, and $B_s$\,--\,$\overline{B}_s$ mixing in presence of RH currents (cf.~the discussion below).

According to the latest fits to the data \cite{Aebischer:2019mlg}, a non-standard contribution in the following $2\sigma$ range is preferred 
\begin{align}
\label{eq:C9minusC10}
\Delta C_9^{bs\mu\mu} &= -\Delta C_{10}^{bs\mu\mu} = [-0.70,\,-0.36]\,,
\end{align}
with the best-fit value $\Delta C_9^{bs\mu\mu} = -\Delta C_{10}^{bs\mu\mu} = -0.52$ improving the fit at the level of $6.5\sigma$
with respect to the SM. For similar global analyses see \cite{Alguero:2019ptt,Alok:2019ufo,Ciuchini:2019usw,Kowalska:2019ley,Arbey:2019duh,Bhattacharya:2019dot,Biswas:2020uaq}.
As discussed above, in presence of scalar mixing, our NP contribution is not exactly of the type $\Delta C_9^{bs\mu\mu} = -\Delta C_{10}^{bs\mu\mu}$. Therefore, in the following we employ a parameterisation of the two-dimensional $(\Delta C_9^{bs\mu\mu},\,\Delta C_{10}^{bs\mu\mu})$ fit result presented in \cite{Aebischer:2019mlg}.\footnote{In practice we require that the values of $\Delta C_9^{bs\mu\mu}$ and $\Delta C_{10}^{bs\mu\mu}$ evaluated in our model are within the 2$\sigma$ likelihood contour of the global fit, as shown in Figure 1 of  \cite{Aebischer:2019mlg}.}

The $SU(2)_L$ counterpart of the left diagram in \fref{fig:bsll} contributes to processes such as $B\to K^{(*)}\nu\bar{\nu}$, 
which can pose a substantial constraint to theories addressing the $b\to s\mu\mu$ anomalies, as pointed out in \cite{Alonso:2015sja,Calibbi:2015kma}. 
However, as we will see below, these bounds are subdominant within our model. The relevant expressions can be found in the Appendix \ref{app:bsnunu}.

The most relevant constraint on the product $\lambda^{Q_d}_3 \lambda^{Q_d\,*}_2$, which enters the NP contribution to 
$\Delta C_9^{bs\mu\mu} = -\Delta C_{10}^{bs\mu\mu}$, is given by $B_s$\,--\,$\overline{B}_s$ oscillations. 
Similarly, in presence of a sizeable $\lambda^Q_1$ coupling, we will have a contribution to $B$\,--\,$\overline{B}$ mixing.
Assuming as above small $\lambda^D_i$ couplings to RH down quarks, our NP will contribute to the following $\Delta B=2$
operators:
\begin{align}
\mathcal{H}^{bd_i}_{\rm eff} \supset  C_1^{bd_i} \,(\overline{d_i}\gamma_\mu P_L b)( \overline{d_i}\gamma^\mu P_L b)  +{\rm h.c.}\,,
\quad\quad {\rm with}~d_i=d,s.
\end{align}
Using the results of~\cite{Arnan:2019uhr,Kawamura:2017ecz}, we find for the contribution of a $Q^\prime-\Phi_S$ box diagram:
\begin{align}
\Delta  C_1^{bd_i} = \frac{(\lambda^{Q_d}_3 \lambda^{Q_d\,*}_{i})^2}{128\pi^2} \sum_{\alpha=1,2}\frac{|U_{1\alpha}|^4}{M_{S_\alpha}^2} 
F\left(\frac{M^2_Q}{M_{S_\alpha}^2}\right),~
\end{align}
where
\begin{align}
F(x)\equiv \frac{x^2-1-2x\log x}{(x-1)^3}.
\end{align}
Real and imaginary parts of these operators are constrained by, respectively, $B_s$\,--\,$\overline{B}_s$ and $B$\,--\,$\overline{B}$
mass differences and CP violation observables. Given that a sizeable value of $\lambda^{Q_d}_1$ would be subject to analogous (but more stringent) constraints from $K$\,--\,$\overline{K}$ mixing (see below), and from the neutron EDM 
(as we will discuss in the following subsection), here we assume $\lambda^{Q_d}_1 \ll \lambda^{Q_d}_2,~\lambda^{Q_d}_3$ and focus on $B_s$\,--\,$\overline{B}_s$ mixing only.
For simplicity, we also assume that the phase of $\lambda^{Q_d}_3 \lambda^{Q_d\,*}_2$ (hence of our contribution to $\Delta  C_1^{bs}$) is suppressed enough so to consider only the bound from the $B_s$\,--\,$\overline{B}_s$ mass difference $\Delta m_s$.
Using the formalism in \cite{Buras:2001mb,Buras:2002vd,Becirevic:2001jj}, and taking the 
%and taking the 2018 UTfit result $\Delta m_s^{\rm SM} = (17.3\pm 0.85)\,{\rm ps}^{-1}$ \cite{Bona:2007vi}, we obtain the following bound:
recent sum-rule and lattice based calculation giving $\Delta m_s^{\rm SM} = (18.4^{+0.7}_{-1.2})\,{\rm ps}^{-1}$ \cite{DiLuzio:2019jyq}, we obtain the following bound:
\begin{align}
\label{eq:Deltams}
\Delta  C_1^{bs} < 2.1 \times 10^{-5}~{\rm TeV}^{-2},
\end{align}
where we take the matching scale of the Wilson coefficient at 1 TeV scale for definiteness. 
The above bound is consistent with that calculated in \cite{Silvestrini:2018dos}. 
%Employing other recent evaluations of $\Delta m_s^{\rm SM}$, we obtain similar results.\footnote{In particular, we find that the most recent sum-rule and lattice based calculation presented in \cite{DiLuzio:2019jyq}, giving $\Delta m_s^{\rm SM} = (18.4^{+0.7}_{-1.2})\,{\rm ps}^{-1}$, translates into $\Delta  C_1^{bs} < 1.7 \times 10^{-5}~{\rm TeV}^{-2}$.}

\begin{figure}[t]
\centering
\includegraphics[width=0.6\textwidth]{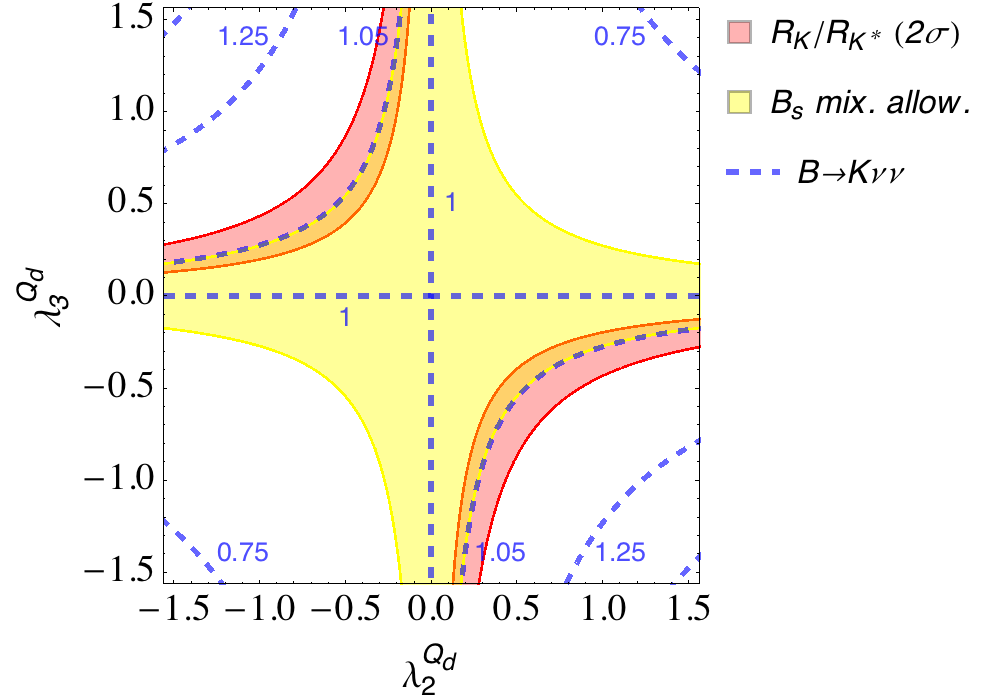}
\caption{Illustrative example on the $\left(\lambda^{Q_d}_2,~\lambda^{Q_d}_3\right)$ plane of the region favoured by $b\to s\ell\ell$ according to \cite{Aebischer:2019mlg} (red) and the one allowed by $B_s$\,--\,$\overline{B}_s$ mixing bound, \eref{eq:Deltams} (yellow). Contours show  $\Gamma(B\to K^{(*)}\nu\bar{\nu})$
normalised by the SM value. The model's parameters were set to: $M_S = 350$ GeV, $M_D = 500$ GeV, $a_H = 20$ GeV, $M_L = 800$ GeV, $M_Q = 1.5$ TeV, $\lambda^L_2 = 1.7$, $\lambda^E_2 = -1.3$. All the other couplings were set to 0. 
 \label{fig:Bobs}}
\end{figure}

The $B_s$\,--\,$\overline{B}_s$ mixing constraint and the $2\sigma$-favoured region for $b\to s\ell\ell$ are shown in \fref{fig:Bobs} for an illustrative choice of the parameters of the model. In this example, the vectorlike quark only couples to 2nd and 3rd generation LH quarks, and the vectorlike lepton  
only couples to muons. The values of the parameters adopted in the above example (especially the large coupling to LH and RH muons) will be better justified in the following subsections.
As we can see, it is possible to find a setup of the parameters for which a good fit of $b\to s\ell\ell$ data is compatible with the
$B_s$\,--\,$\overline{B}_s$ mixing bound, although such a constraint is particularly severe. An improvement of the theoretical determination of $\Delta m_s^{\rm SM}$ would allow therefore to test an explanation of this kind of the $B$-physics anomalies. This is a common feature of models addressing 
 $b\to s\ell\ell$ at one loop (cf.~for instance the discussion in \cite{Arnan:2016cpy}). The figure also shows that in the (orange) region where $b\to s\ell\ell$  and $B_s$\,--\,$\overline{B}_s$ are compatible the rate of $B\to K^{(*)}\nu\bar{\nu}$ (cf.~Appendix \ref{app:bsnunu} for the relevant expressions) deviates from the SM prediction by at most 5\% and thus does not further constrain the model at present \cite{Buras:2014fpa}.   

Concerning the possible couplings to RH down quarks, cf.~\eref{eq:lag}, we notice that non-vanishing $\lambda^D_{2,3}$ would generate other operators contributing to $B_s$\,--\,$\overline{B}_s$, including the LR and RR currents listed in the Appendix \ref{app:DeltaM}. 
The coefficient of LR operators are subject to a bound that is about a factor 3 stronger than the one given above for $\Delta  C_1^{bs}$ \cite{Silvestrini:2018dos}. 
Moreover, dipole operators and scalar operators would arise giving substantial contributions to respectively $b\to s\gamma$ and $B_s\to\mu\mu$,
cf.~Appendix \ref{app:bsll} and \ref{app:bsg}.\footnote{Using expressions and bounds reported in the Appendix (cf.~Eqs.~(\ref{eq:Bs},\,\ref{eq:bsg}))
we checked that, despite the helicity-enahanced scalar contributions, $\lambda^D_{2,3}\sim\mathcal{O}(1)$ are still compatible with the measured rate of $B_s\to\mu\mu$, whereas $b\to s\gamma$ sets a substantial constraint ($|\lambda^D_{2}| \lesssim 0.01$ for our benchmark spectrum).}
Considering that the improvement to the $b\to s\ell\ell$ fit in presence of RH currents is not dramatic \cite{Alguero:2019ptt,Alok:2019ufo,Ciuchini:2019usw,Aebischer:2019mlg,Kowalska:2019ley,Arbey:2019duh,Bhattacharya:2019dot}, the compatibility between $b\to s\ell\ell$ and other $b-s$ transitions prefers that the couplings $\lambda^D_{2,3}$ are suitably suppressed. Here and in the following, we just set them to zero for simplicity.

Finally, we discuss bounds from  $K$\,--\,$\overline{K}$ mixing.  Using the limits on Wilson coefficients reported in \cite{Isidori:2013ez}, we find that for the benchmark point of \fref{fig:Bobs} and a real $\lambda^{Q_d}_2$ the bounds from $\Delta m_K$ and $\epsilon_K$ translate to, respectively, $|\lambda^{Q_d}_1|\lesssim 0.056/|\lambda^{Q_d}_2|$ and $|(\lambda^{Q_d}_1)|\times \sqrt{\arg{(\lambda^{Q_d}_1)}} \lesssim 2.4\times 10^{-3}/|\lambda^{Q_d}_2|$. Notice that, if we use the basis of \eref{eq:lambdaQ} (which is aligned to the up sector) with a vanishing $\lambda^Q_1$, $\lambda^{Q_d}_1$ is still generated by CKM rotations, which sets a bound in particular on $\lambda^Q_2$. We find that the kaon bounds are then fulfilled for $|\lambda^Q_2| \lesssim 0.15$. This limit is unavoidable since we indeed need a flavour structure well aligned to the up sector, 
because $\lambda^{Q_u}_1 \ll 1$ is required to fulfil the neutron EDM bound, as we will see below.
Hence, \fref{fig:Bobs} tells us the our $B$-physics observables can be fitted if  $|\lambda^{Q_d}_3|\approx|\lambda^Q_3| \gtrsim 1$.

\subsection{CP violation in charm decays}
\label{sec:DeltaACP}
The LHCb experiment has recently established CP violation in the charm sector, by measuring the difference 
of the time-integrated CP asymmetries in the $|\Delta C| =1$ decays $D^0 \to K^+ K^-$ and $D^0 \to \pi^+ \pi^-$ \cite{Aaij:2019kcg}:
\begin{align}
\Delta A_{\rm CP} \equiv A(K^+K^-) - A(\pi^+\pi^-) = (-15.4\pm2.9)\times 10^{-4}\,,
\label{eq:dacp-exp}
\end{align}
where
\begin{align}
A(f) \equiv \frac{\Gamma({D}^0\to f)-\Gamma(\overline{D}^0\to f)}{\Gamma({D}^0\to f)+\Gamma(\overline{D}^0\to f)},\quad\quad f=K^+K^-,\,\pi^+\pi^-.
\end{align}
This observable is mostly sensitive to direct CP violation \cite{Grossman:2006jg}.

Interpreting the LHCb result is not straightforward, given the notorious difficulty of performing calculations at the charm mass scale. 
In the SM one gets $\Delta A^{\rm SM}_{\rm CP}\approx -0.0013\times  \Im(\Delta R^{\rm SM})$ (see e.g.~\cite{Isidori:2011qw,Giudice:2012qq}) 
where $\Delta R^{\rm SM}$ encodes ratios of hadronic amplitudes naively expected to be of the order $\Delta R^{\rm SM} \approx \alpha_s(m_c)/\pi \approx 0.1$. This estimate is supported by the recent calculation in \cite{Chala:2019fdb} giving $|\Delta A^{\rm SM}_{\rm CP}|\le 3\times 10^{-4}$,
and implies a substantial discrepancy with the measured value. However, it is not possible to exclude that large non-perturbative 
effects in  $\Delta R^{\rm SM}$ could enhance the SM prediction up to the value observed by LHCb \cite{Brod:2011re,Brod:2012ud,Li:2012cfa,Grossman:2019xcj,Cheng:2019ggx,Li:2019hho}. Here, we are going to speculate about the possible NP origin of $\Delta A_{\rm CP}$. For the implications on other NP models of large CP violation in charm decays see also \cite{Dery:2019ysp,Chala:2019fdb,Giudice:2012qq,Calibbi:2013mka}.

Possible NP effects in $\Delta A_{\rm CP}$ are encoded in the $|\Delta C| =1$ effective Hamiltonian:
\begin{align}
\mathcal{H}^{cu}_{\rm eff } = -\frac{4G_F}{\sqrt{2}} \sum_i C_i^{cu} Q_i^{cu} +{\rm h.c.}\,. 
\end{align}
The full list of operators can be found in \cite{Isidori:2011qw}. 
Following \cite{Isidori:2011qw,Giudice:2012qq}, here we are interested in the NP contribution to the chromo-magnetic dipole operators 
that can give rise to sizeable $|\Delta C| =1$ effects without inducing unacceptably large contribution to $|\Delta C| =2$ operators, i.e.~to $D$\,--\,$\bar{D}$ mixing. These operators read:
\begin{align}
 Q_8^{cu}  =\frac{m_c}{16\pi^2} g_s\, \overline{u} \sigma_{\mu\nu} T^a G_a^{\mu\nu} P_R c\,, \quad\quad
  \widetilde{Q}_8^{cu} =\frac{m_c}{16\pi^2} g_s\, \overline{u} \sigma_{\mu\nu} T^a G_a^{\mu\nu} P_L c\,. 
\end{align}
The resulting $\Delta A_{\rm CP}$ is  \cite{Isidori:2011qw,Giudice:2012qq}
\begin{align}
\Delta A_{\rm CP} &\approx - \frac{2}{\sin\theta_c} \left[\Im(V_{cb}^*V_{ub}) \Im(\Delta R^{\rm SM}) -9\sum_i  \Im\left(\Delta C_i^{cu}(m_c)\right) \Im(\Delta R^{\rm NP}_i)\right],
\label{eq:dacp}
\end{align}
where $\Delta R^{\rm SM}$ and $\Delta R^{\rm NP}_i$ are combinations of hadronic amplitudes, and $\Delta C_i^{cu}(m_c)$ are the NP contributions to 
the coefficients of $Q_8^{cu}$ and  $ \widetilde{Q}_8^{cu}$ at the $m_c$ scale.
\begin{figure}[t]
\centering
\includegraphics[width=0.48\textwidth]{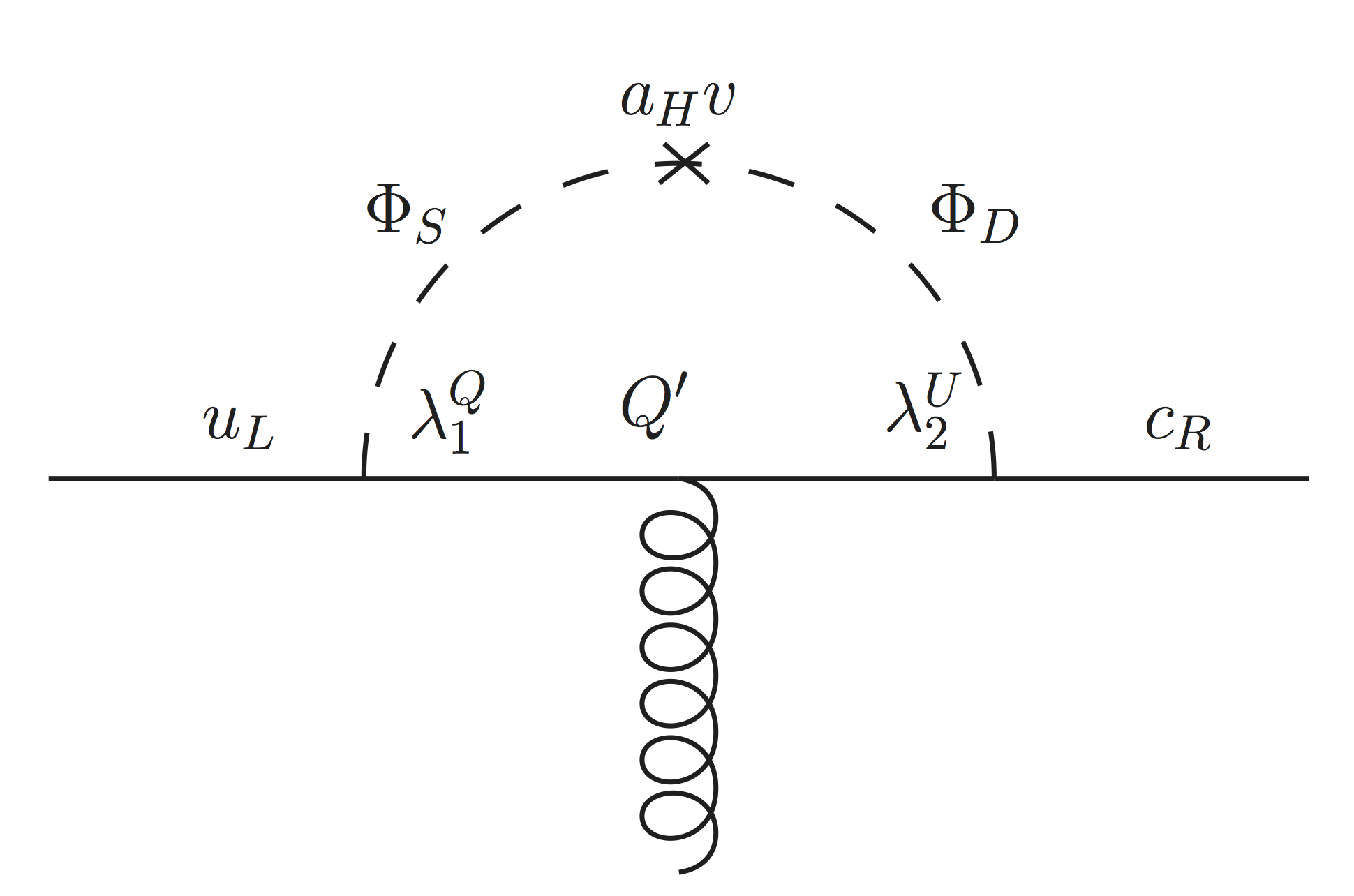}
\hfill
\includegraphics[width=0.48\textwidth]{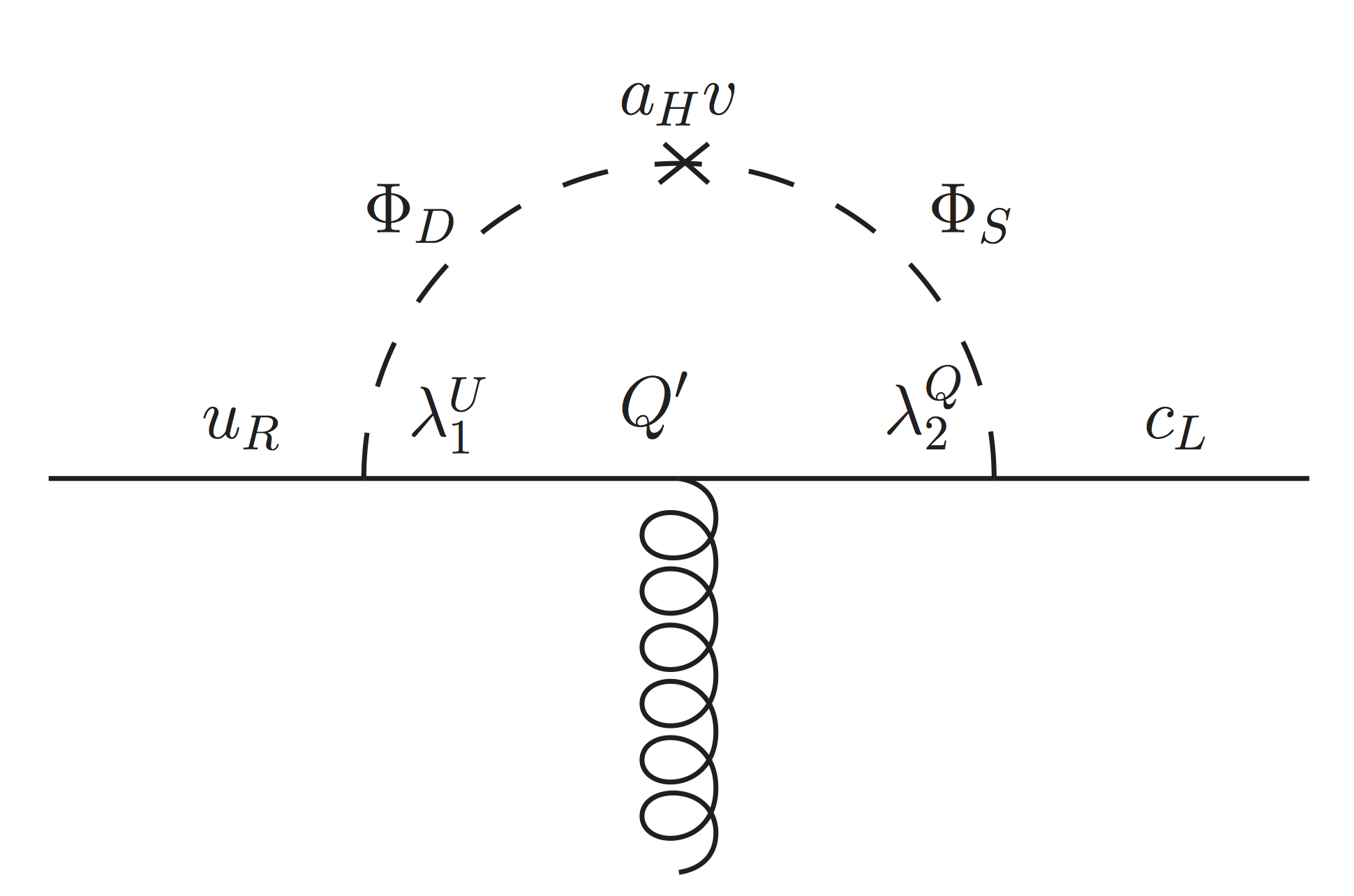}
\caption{Diagrams generating $|\Delta C| =1$ chromo-magnetic dipole operators.\label{fig:C8}}
\end{figure}

In our model the chromo-magnetic operators are generated by the two diagrams shown in \fref{fig:C8}. 
Adapting again the general formulae derived in~\cite{Arnan:2019uhr}, we obtain:
\begin{align}
\label{eq:C8}
\Delta C^{cu}_8(\Lambda) &\simeq  \frac{M_Q}{\sqrt{2} G_F m_c} \lambda^U_{2} \lambda^{Q\,*}_{1}\sum_{\alpha=1,2}   \frac{U_{2\alpha} U^*_{1\alpha}  }{M_{S_\alpha}^2}\, G_8\left(\frac{M^2_Q}{M^2_{S_\alpha}} \right), \\
\label{eq:C8t}
\Delta \widetilde{C}^{cu}_8(\Lambda) &\simeq  \frac{M_Q}{\sqrt{2} G_F m_c} \lambda^{Q}_{2} \lambda^{U*}_{1}\sum_{\alpha=1,2}   \frac{U_{1\alpha} U^*_{2\alpha}  }{M_{S_\alpha}^2}\, G_8\left(\frac{M^2_Q}{M^2_{S_\alpha}} \right), 
\end{align}
where we only show the dominant chirally-enhanced LR contributions (the subdominant LL and RR terms can be found in \cite{Arnan:2019uhr}),
$\Lambda$ is the matching scale ($\approx$ 1 TeV) and the loop function reads:
\begin{align}
G_8(x)  \equiv \frac{x^2 -2 x\log x-1}{8 (x-1)^3} \, .
\end{align}
The evolution of the coefficients down to $m_c$ can be computed with the standard formulae for the QCD running (see \cite{Buchalla:1995vs})
and numerically gives $\Delta C^{cu}_8(m_c) \simeq 0.41 \times\Delta C^{cu}_8(1\,{\rm TeV})$.

In the following we are assuming that the dominant contribution to $\Delta A_{\rm CP}$ is due to new physics,
and we are employing the estimate for $\Delta R^{\rm NP}_{8,\widetilde{8}}$ given in~\cite{Isidori:2011qw,Giudice:2012qq}:
\begin{align}
\left|\Im (\Delta R^{\rm NP}_{8,\widetilde{8}})\right|\approx 0.2,
\end{align}
which was obtained using naive factorisation and assuming maximal strong phases.
Under the above assumptions (which are subject to $\ord{1}$ uncertainties), the value of  $\Delta A_{\rm CP}$  measured by LHCb is saturated for  	
\begin{align}
\label{eq:DACP-range}
\left|\Im \left(\Delta C^{cu}_{8}(m_c) + \Delta \widetilde{C}^{cu}_{8}(m_c)\right)\right| \approx  (5 \div 12)\times 10^{-4},
\end{align}
where we considered the 2$\sigma$ range of \eref{eq:dacp-exp}.
\begin{figure}[t]
\centering
\includegraphics[width=0.6\textwidth]{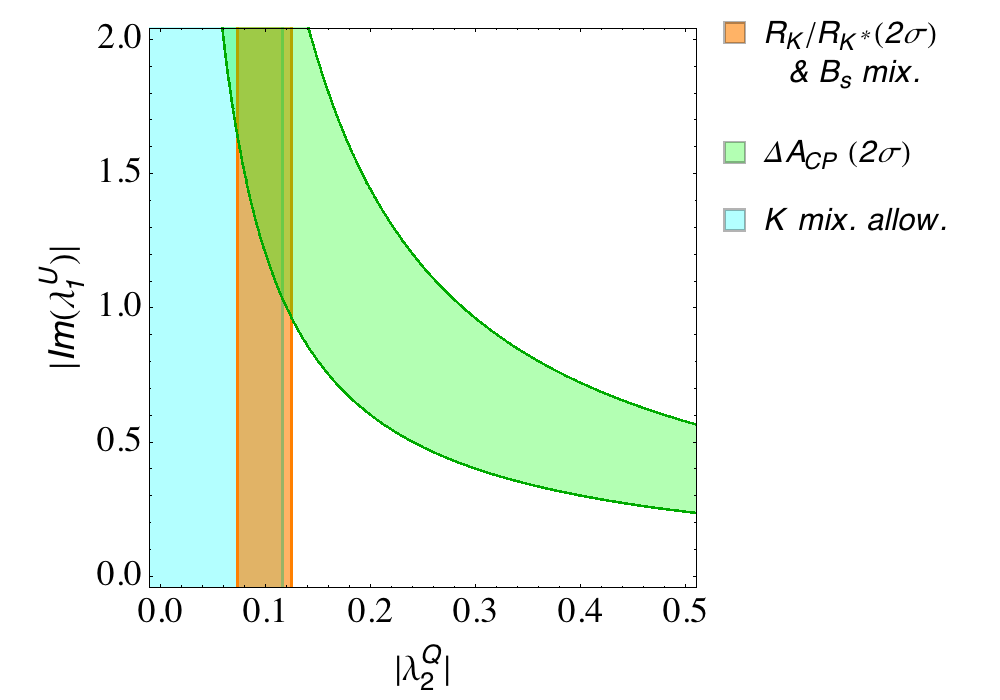}
\caption{Regions favoured by $\Delta A_{\rm CP}$ (green), $b\to s\ell\ell$ (red), and allowed by $B_s$\,--\,$\overline{B}_s$ (yellow) and $K$\,--\,$\overline{K}$ (cyan) on the $\left(\lambda^{Q}_{2},~\Im(\lambda^{U}_{1})\right)$ plane, for $\lambda^{Q}_{3}=-1.5$. The other parameters were set as in \fref{fig:Bobs}. \label{fig:DeltaACP}}
\end{figure}

By inspecting Eqs.~(\ref{eq:C8},\,\ref{eq:C8t}), we can see that a large effect in $\Delta A_{\rm CP}$ is more easily induced by $\Delta \widetilde{C}^{cu}_8$. The reason is that a sizeable value of $\lambda^Q_{2}$ is required by fitting the $b\to s \ell\ell$ data, so that $ \lambda^{Q}_{1}$ is tightly constrained
by $K$\,--\,$\bar{K}$ mixing, as discussed in the previous subsection. On the other hand, a complex $\lambda^{U}_{1}$ can easily account for the observed value of $\Delta A_{\rm CP}$. 
However, a sizebale (complex) value of $\lambda^{U}_{1}$ can contribute to the up quark (and thus to the neutron) EDM, via the flavour-conserving counterpart of the diagrams in \fref{fig:C8}. This sets a further constraint on $\lambda^{Q}_{1}$. Employing the formalism presented in the Appendix \ref{app:dn}, we find $|\lambda^{Q_u}_1| \lesssim 0.5\times10^{-3}/|\lambda^{U}_{1}|$. In order to evade this stringent bound, we can set $\lambda^{Q_u}_1=\lambda^{Q}_1 \approx 0$. This kind of up-sector alignment however implies that $\lambda^{Q_d}_1$ can not be arbitrarily small, so that kaon sector constraints
set an upper bound on $\lambda^Q_2$, as discussed at the end of \sref{sec:Bphys}.
In \fref{fig:DeltaACP}, we show that, for our benchmark point, an overlap between the regions favoured by $\Delta A_{\rm CP}$ (according to \eref{eq:DACP-range}) and $b\to s\ell\ell$ is obtained for $|\Im(\lambda^{U}_{1})|\gtrsim 1$, while fulfilling at the same time bounds from $K$\,--\,$\bar{K}$ mixing. Such large contribution to $\Delta A_{\rm CP}$ is achieved thanks to the chiral enhancement that follows from the singlet-doublet mixing (which is a peculiarity of this model), as shown in \fref{fig:C8}. 

Let us finally discuss other possible constraints from the up-type quark sector. As pointed out in \cite{Isidori:2011qw}, the $D$\,--\,$\bar{D}$ mixing constraint is irrelevant when the above dipole contributions saturate the observed $\Delta A_{\rm CP}$ value. By employing the expressions reported in the Appendix \ref{app:DeltaM}, we have explicitly checked that this is indeed the case: the limit on $\Delta  C_2^{cu}$ reported in \cite{Isidori:2011qw} translates into
the irrelevant bound $|\lambda^{U}_{1}|\lesssim 1/|\lambda^{Q}_{2}|$ for the benchmark point of \fref{fig:DeltaACP}.
Furthermore, a constraint on the product $\Im\left(\lambda^{U}_{2}\lambda^{Q\,*}_{2}\right)$ arises from a limit on the charm chromo-EDM that was obtained in \cite{Sala:2013osa} considering its contribution to the neutron EDM. This requires that $\lambda^{U}_{2}$ is either small or to large extent real: for a real $\lambda^{Q}_{2}$, we get $\Im(\lambda^{U}_{2})\lesssim 10^{-2} (0.15/|\lambda^{Q}_{2}|)$.

\subsection{Muon $g-2$}
\begin{figure}[t]
\centering
\includegraphics[width=0.48\textwidth]{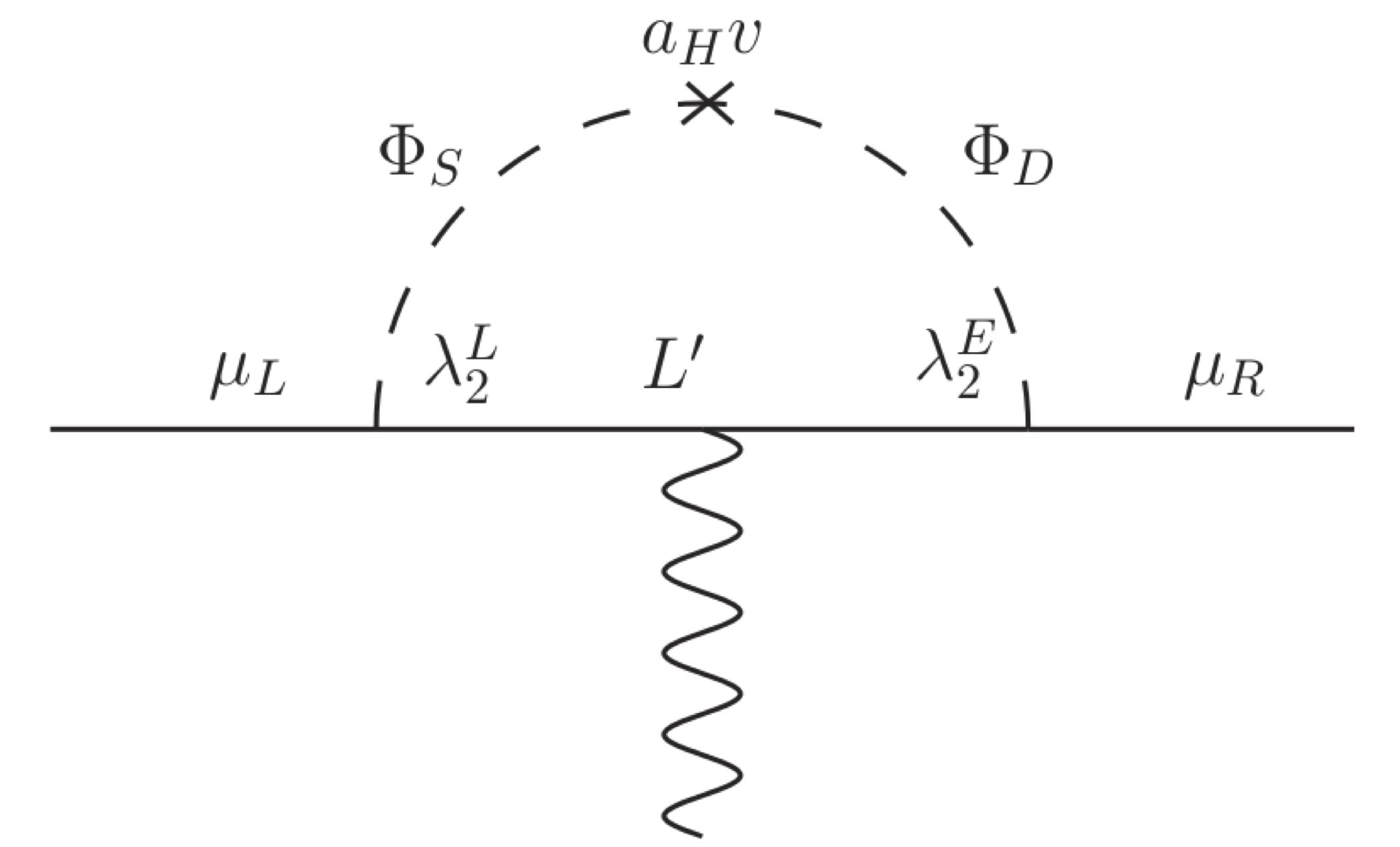}
\caption{Leading contribution to the muon $g-2$.\label{fig:g-2}}
\end{figure}
Loop diagrams involving the extra scalars and the vectorlike lepton $L^\prime$ contribute to
the anomalous magnetic moment of the muon, $a_\mu \equiv (g-2)_\mu$. For a recent review see \cite{Lindner:2016bgg}.
According to the classification of~\cite{Calibbi:2018rzv}, this is a `scalar LR' (SLR) model, namely one that yields a chirally-enhanced
contribution\footnote{This means a contribution for which the chirality flip required by the dipole transition is realised through a Higgs vev insertion inside the loop, instead of on an external muon line, thus avoiding the suppression proportional to the small muon Yukawa coupling. See \cite{Calibbi:2018rzv} for a thorough discussion.} to $a_\mu$, as a consequence of the scalar singlet-doublet mixing. 
The relevant diagram is depicted in \fref{fig:g-2}. 
The leading chirally-enhanced contribution to $a_\mu$ reads:
\begin{align}
\label{eq:damu}
\Delta a_\mu \approx ~& - \frac{m_\mu M_L}{8 \pi^2 } \sum_{\alpha=1,2}   \frac{\Re(\lambda^L_{2} \lambda^{E\,*}_{2}\, U_{1\alpha} U^*_{2\alpha})  }{M_{S \alpha}^2}f_{LR} \left( \frac{M_{L}^2}{M_{S_\alpha}^2}\right) ,
\end{align}
where 
\begin{align}
f_{LR} (x)  \equiv  \frac{3-4x+x^2+2 \log x}{2 (x-1)^3} \, .
\end{align}
The expression of the subdominant terms, which are suppressed by a factor $\sim y_\mu$ relative to the above contribution, can be found in~\cite{Calibbi:2018rzv}.

If confirmed, the present discrepancy between the measurement and the SM prediction of $a_\mu$ would require at 1$\sigma$ \cite{Bennett:2006fi,Davier:2010nc,Blum:2013xva}:
\begin{align}
\label{eq:g-2exp}
\Delta a_\mu = a_\mu^{\rm EXP} - a_\mu^{\rm SM} = (2.87 \pm 0.80) \times 10^{-9} \,.
\end{align}

\begin{figure}[t]
\centering
\includegraphics[width=0.6\textwidth]{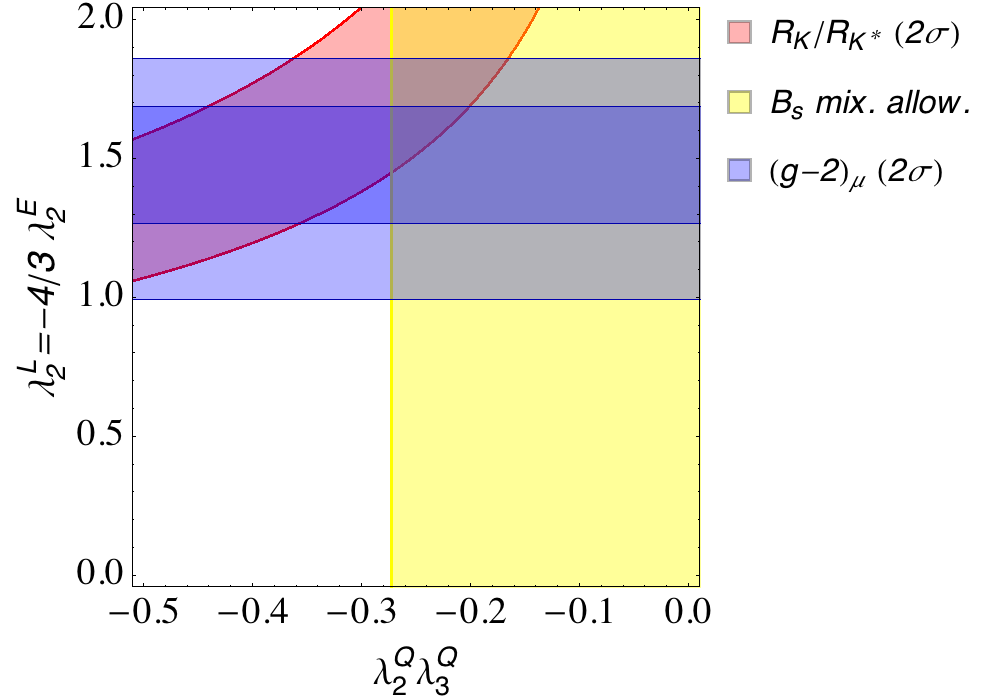}
\caption{Regions favoured by the muon $g-2$ at 2$\sigma$ (blue), $b\to s\ell\ell$ (red), and allowed by $B_s$\,--\,$\overline{B}_s$ (yellow) on the $\left(\lambda^{Q}_{2}\lambda^{Q}_{3},~\lambda^{L}_{2}=-4/3\, \lambda^{E}_{2}\right)$ plane.
The other parameters were set as in \fref{fig:Bobs}.  \label{fig:gm2}
}
\end{figure}

In \fref{fig:gm2}, we show the value of $\lambda^{L}_{2}$ (setting  $\lambda^{E}_{2}=-0.75\,\lambda^{L}_{2}$) for which \eref{eq:damu} provides a (positive) contribution $\Delta a_\mu$ of the size required by the above discrepancy  (within $2\sigma$). As we can see, for large enough couplings to the LH and RH muons, such region overlaps to that favoured by the 
$b\to s\mu\mu$ anomalies (and still allowed by the $B_s$\,--\,$\overline{B}_s$ mixing constraint). The figure shows that a large value of the coupling $\lambda^L_2$ ($\approx 1.5$ for this numerical example) is needed such that the contribution $\Delta C_9^{bs\mu\mu} = -\Delta C_{10}^{bs\mu\mu}$ (that depends on $\lambda^E_2$ only via the singlet-doublet mixing, hence mildly) accounts for the $b\to s\mu\mu$ anomalies and the $B_s$\,--\,$\overline{B}_s$ 
bound is simultaneously evaded. Given the chirally-enhanced contribution of \eref{eq:damu},  accommodating the observed value of $a_\mu$ and  
$b\to s\mu\mu$ simultaneously thus requires either sizeable singlet-doublet mixing and small $\lambda^E_2$, or small mixing and large $\lambda^E_2$ (as in the example shown in \fref{fig:gm2}).

Sizeable couplings to electron or tau ($\lambda^{L,E}_1,~\lambda^{L,E}_3$) would induce lepton-flavour-violating (LFV) dipole operators through diagrams similar to \fref{fig:g-2}, thus being subject to the tight constraints from searches of LFV decays such as $\mu\to e\gamma$ and $\tau\to \mu\gamma$ (for a recent overview, cf.~\cite{Calibbi:2017uvl}). As discussed in \cite{Calibbi:2018rzv}, one indeed finds that the current limits on LFV processes \cite{TheMEG:2016wtm,Aubert:2009ag} and the central value in \eref{eq:g-2exp} imply:
\begin{align}
\left|\lambda^{L\,,E}_1/\lambda^{L\,,E}_2\right| \lesssim 1.8\times 10^{-5} \left( \frac{\Delta a_\mu}{2.87\times 10^{-9}} \right),\quad
\left|\lambda^{L\,,E}_3/\lambda^{L\,,E}_2\right| \lesssim 1.4\times 10^{-2} \left( \frac{\Delta a_\mu}{2.87\times 10^{-9}} \right).
\end{align}
The stringent limit on the electron couplings, in particular, does prevent any sizeable contribution to the $g-2$ of the electron.
This observable also exhibits a mild tension with the SM prediction: $\Delta a_e = a_e^{\rm EXP} - a_e^{\rm SM} = -(0.88 \pm 0.36) \times 10^{-12}$ \cite{Parker:2018vye}.
In order to account for that, one would then need to extend our model, e.g.~introducing multiple generations of vectorlike leptons (coupling either to electrons or to muons, not to both) along the lines of the models discussed in \cite{Crivellin:2018qmi}.
		
\subsection{Summary: Size and flavour structure of the couplings}
\label{sec:couplings}
We conclude this section by summarising the structure of the couplings of our new particles that we can infer from the above discussion.
For a TeV-scale spectrum	of the new fields $Q^\prime$, $L^\prime$, $\Phi_S$, $\Phi_D$ and a moderate singlet-doublet mixing (as in the illustrative example adopted in Figs.~\ref{fig:Bobs},~\ref{fig:DeltaACP},~\ref{fig:gm2}), the model can successfully account for $b\to s\mu\mu$, $\Delta A_{CP}$, muon $g-2$ if the following minimal set of ingredients is present:
\begin{itemize}
\item Sizeable couplings of the vectorlike quark $Q^\prime$ to LH bottom and strange quarks (with opposite signs): $\lambda^Q_2 \lambda^Q_3 \approx -(0.4\div0.5)^2$, cf.~Figs.~\ref{fig:Bobs},\,\ref{fig:DeltaACP},\,\ref{fig:gm2}; 
\item $\mathcal{O}(1)$ couplings of  $L^\prime$ to LH and RH muons, for the sake of the combined explanation of 
$b\to s\mu\mu$ and $\Delta a_\mu$: $|\lambda^L_2|\gtrsim 1.5$, $|\lambda^E_2|\gtrsim 1$,  cf.~Fig.~\ref{fig:gm2}; 
\item A substantial (complex) coupling of $Q^\prime$ to the RH up quark, in order to induce large CP violation in charm decays:
$|\Im(\lambda^U_1)|\gtrsim 1$,  cf.~Fig~\ref{fig:DeltaACP};
\item Suppressed coupling to LH up and down quarks due to bound from the neutron EDM and $K$\,--\,$\overline{K}$ mixing: 
to satisfy both at the same time, we need $|\lambda^Q_1| \lesssim 10^{-3}$,  $|\lambda^Q_2| \lesssim 0.15$;
\item Small to mildly-suppressed couplings of $Q^\prime$ to RH down-type quarks: $|\lambda^D_1| \lesssim 10^{-(1\div2)}$ (from $K$\,--\,$\overline{K}$ mixing), $|\lambda^D_{2}| \lesssim 10^{-2}$ (mainly from $b\to s\gamma$),
$|\lambda^D_{3}| < |\lambda^Q_{3}|$ (to prevent large RH-current effects in $b\to s\mu\mu$ and $B_s$\,--\,$\overline{B}_s$ mixing);
\item Very small couplings to LH and RH electron and tau, in order to evade bounds from LFV processes: 
$|\lambda^{L\,,E}_1| \lesssim 10^{-5}$, $|\lambda^{L\,,E}_3| \lesssim 10^{-2}$.
\end{itemize}
Although the above pattern is not generic, it is certainly conceivable, especially if enforced by a flavour symmetry. 
In particular, notice that the couplings to quarks are in principle compatible with a SM-like hierarchical structure: $\lambda^Q_1 \ll \lambda^Q_2 \lesssim \lambda^Q_3$, $\lambda^D_1 < \lambda^D_2 < \lambda^D_3$, $\lambda^U_1 \lesssim \lambda^U_2 \lesssim \lambda^U_3$ (the absolute values of the couplings to RH charm and top being virtually unconstrained).

As we can see from the above summary, some of the couplings $\lambda_X$ may be required to be larger than one, $\lambda_X \gtrsim 1$, for the sake of a combined explanation of the flavour effects under discussion. These large values can cause a Landau pole at low energies, which limits the range of scales, for which our simplified model is consistent. As usual, a Landau pole would indicate the scale at which the model needs to be embedded in a more fundamental theory.
In order to estimate such a scale, we employ one-loop renormalization group equations:
\begin{align}
\frac{d\lambda_X}{dt}\simeq \frac{1}{16\pi^2}C_X \lambda_X^3,
\end{align}
where $t\equiv \ln\left( \mu/M\right)$ (with $M\approx 1$ TeV being the typical mass scale of our extra fields), $C_X = 5/2$ for $\lambda_X=\lambda^{L,E}_i$,  $C_X = 9/2$ for 
$\lambda_X = \lambda^{Q,U,D}_i$, and we neglected terms proportional to the gauge couplings, which are subdominant in the regime $\lambda_X \gtrsim 1$, as well as terms proportional to Yukawa couplings of another kind, $\sim\lambda_X^{\prime\,2}$.\footnote{This latter choice follows from the assumption that there is only one $\ord{1}$ Yukawa for each single fermion flavour, which is consistent with the flavour constraints discussed above.} Notice that such terms tend to slow down the running, hence they would just relax the bounds discussed in the following. From the above equation, one obtains that the couplings remain finite for scales $\mu\lesssim \mu_\text{max}$, where
\begin{align}
\mu_\text{max} = M e^{8\pi^2/C_X\lambda^2_X(M)}\,.
\end{align}	
Thus we find that $\mu_\text{max}/M \approx (5\times 10^{13},\,10^6,\,3000)$ for respectively $|\lambda_i^{L,E}(M)| = (1,\,1.5,\,2)$, and 
$\mu_\text{max}/M \approx (4\times 10^{7},\,3000,\,80)$ for $|\lambda_i^{Q,U,D}(M)| = (1,\,1.5,\,2)$. 
Therefore the couplings to the vectorlike quark set the most stringent constraint on the scale to which the model can be safely extrapolated. If we choose any coupling $|\lambda_i^{Q,U,D}(M)| \approx 2$, such scale is $\approx 100$ TeV.

In the following section, we are going to assume the pattern of couplings summarised above, and discuss in better detail the new particle spectrum 
selected by the flavour anomalies and its consequences for LHC and dark matter.

\section{Combined constraints, spectrum, and dark matter}
\label{sec:DM-LHC}

\subsection{LHC phenomenology}
\label{sec:LHC}

The new states of our model can only be produced in pairs at colliders, as a consequence of the $\mathbf Z_2$ symmetry. For the same reason, they undergo decays ending in a SM particle plus the lightest $\mathbf Z_2$-odd particle, which we assume to be the lightest neutral scalar $S_1$, in order to address the DM problem (see \ref{sec:DM} for further details). All these features remind of supersymmetric models and, likewise, collider signatures will include energetic jets or leptons plus missing transverse momentum $\met$. These are exactly the same final states predicted in R-parity conserving supersymmetric theories in the case of production of squarks and sleptons, which subsequently decay to a lighter invisible neutralino. Searches for supersymmetry at the LHC can be thus used to set limits on the masses of our new particles too. A detailed study of the bounds on the different production modes and decay chains would be beyond the scope of this work. Here we focus on a number of simplified topologies, in order to demonstrate that large regions of the parameter space that are relevant for the flavour processes discussed in the previous section are not excluded by current LHC searches (but are possibly in the reach of future LHC runs). In particular, we consider:
\begin{enumerate}	
\item[\bf 1.] Strong production of the $Q^\prime$ states, i.e. $pp\to U^\prime \overline{U}^\prime$ and $pp\to D^\prime \overline{D}^\prime$;
\item[\bf 2.] Electroweak (Drell-Yan) production of $L^{\prime}$: $pp\to L^{\prime\,+} {L}^{\prime -}$, $pp\to L^{\prime\,0} {L}^{\prime 0}$;
\item[\bf 3.] Electroweak production of the scalar (mostly) doublet states, i.e.~(assuming $M_S < M_D$):  $pp\to S^+S^-$, $pp\to S^\pm S_2$, etc. 
\end{enumerate}	
The decays of these particles can be visualised in the sketch of \fref{fig:spectrum}. In the following, we are going to discuss them in turn, together with
the resulting LHC signatures and searches. 

%\vspace{-0.3cm}
\paragraph{1. $Q^\prime$  production.} 
Given the pattern of the couplings discussed in \sref{sec:couplings}, the $Q^\prime$ states will mostly decay through $\lambda^Q_3$ to top and bottom  (leading to $U^\prime \to t \,S_1$ and $D^\prime \to b\, S_1$), and through $\lambda^Q_2$ to strange and charm (thus giving $U^\prime \to j \,S_1$ and $D^\prime \to j\, S_1$). Rates of decays into mostly doublet scalar states such as $S_2$ are suppressed as they require singlet-doublet mixing. Furthermore, the decays controlled by $\lambda^U_1$ are typically subdominant since $\Delta A_{CP}$ prefers a moderate value of this coupling. They would anyway lead to more complicated (and possibly phase-space suppressed) decay chains, such as $U^\prime \to j \,S_2 \to j\,h\,S_1$, which are arguably less clean than the above signatures.
A recent analysis performed by the CMS collaboration \cite{Sirunyan:2019ctn}, which employs the full data set of the 13 TeV run, addresses the signatures relevant for this production mode  and the direct decays into $S_1$ discussed above: $t\bar{t}+\met$, $2b$-jets\,+\,$\met$, and $2j+\met$. This search sets a limit on the production cross section of stops, sbottoms, and (a single generation of) squarks that is approximately $\sigma \lesssim 1.7$ fb. Given that for states above $\approx 1$ TeV decaying to much lighter particles the efficiency times acceptance of the search is virtually constant, we can directly translate this limit into a bound on the mass of the $Q^\prime$ fermions (valid if $M_Q\gg M_{S_1}$): $M_Q \gtrsim 1.5$ TeV.\footnote{In order to get this, we employed the production cross section as calculated at LO by means of MadGraph5 \cite{Alwall:2014hca} and we rescaled it by a k-factor of 1.44 obtained by comparing LO and NLO-NLL squark production cross sections \cite{Borschensky:2014cia,Beenakker:2016lwe}. Furthermore, notice that, being the limit reported in \cite{Sirunyan:2019ctn} basically the same for stops, sbottom, and squarks, our estimate does not strongly depend on the branching fractions of $U^\prime(D^\prime) \to t(b) \,S_1$ and $U^\prime(D^\prime) \to j \,S_1$ (controlled by $\lambda^Q_3/\lambda^Q_2$) .} For simplicity, in the previous section, as well as in the following discussion, we set $M_Q = 1.5$ TeV, a value that should be still borderline viable according to the above estimate. We have to keep in mind though that strong production of $Q^\prime$ could be a way to test our scenario at future LHC runs.

%\vspace{-0.3cm}
\paragraph{2. $L^\prime$  production.} 
The charged states can decay directly into muons and $S_1$ through the coupling $\lambda^L_2$, $L^{\prime\,\pm}\to \mu^\pm \,S_1$, while 
the coupling to the scalar doublet $\lambda^E_2$ would induce longer decay chains, e.g.~$L^{\prime\,\pm}\to \mu^\pm \,S_2 \to \mu^\pm \,S_1\, h/Z$.
Similarly, $L^{\prime\,0}$ decays as $L^{\prime\,0}\to \mu^\pm \,S^\mp \to \mu^\pm \,S_1 \,W^\mp$ due to $\lambda^E_2$, while the decay induced by 
$\lambda^L_2$ is completely invisible: $L^{\prime\,0}\to \nu \,S_1$. In the following we are focusing on the simplest topology $pp\to L^{\prime\,+}L^{\prime\,-} \to \mu^+\mu^-+\met$. The latest available analysis of this signal has been presented by ATLAS in \cite{Aad:2019vnb} (see also the results with a smaller data set in \cite{Aaboud:2018jiw}). The resulting bound on the production cross section can be as strong as $\sigma \lesssim 0.2$ fb (for $M_L \gg M_{S_1}$), which corresponds to $M_L \gtrsim 900$ GeV (according to the LO production cross section as calculated by MadGraph5 \cite{Alwall:2014hca}).
This limit is slightly above the benchmark value of $M_L$ employed in the last section, but it is also likely too tight, as the longer decay chain induced by $\lambda^E_2$ would partially dilute the signal and lead to other signatures, which are possibly more challenging to constrain at the LHC (at least if the mass difference between the vectorlike lepton and the scalar doublet is not very large). A more quantitative discussion of the bound on $L^{\prime\,\pm}$ will be presented in \sref{sec:discussion}.

%\vspace{-0.3cm}
\paragraph{3. Scalar doublet production.} 
The production of the states of the scalar doublet, decaying to SM bosons and DM, leads to topologies similar to those sought for in the case of electroweak production of supersymmetric charginos and neutralinos: $pp\to S^+S^- \to W^+W^- +\met$, $pp\to S^\pm S_2 \to W^\pm\, h/Z +\met$. The most sensitive signature is thus again 2 \cite{Aad:2019vnb} or 3 \cite{Sirunyan:2018ubx} leptons 
(from the leptonic decays of the gauge bosons) and $\met$. This searches can constrain Higgsino-like charginos and neutralinos with masses up to about 600 GeV. However, the production cross section for our scalars is much smaller than for a fermion doublet of the same mass. As a consequence,  we can estimate that searches as in~\cite{Sirunyan:2018ubx} are at most sensitive to doublet masses up to 200-300 GeV for a very light singlet-like $S_1$, $M_{S_1} < 100$ GeV. Therefore, as it will be clear from the plots presented in \sref{sec:discussion}, these modes do not represent yet a relevant constraint of the region of the parameters space selected by the flavour observables.

\subsection{Dark matter phenomenology}
\label{sec:DM}
\begin{figure}[t]
\centering
\includegraphics[width=0.85\textwidth]{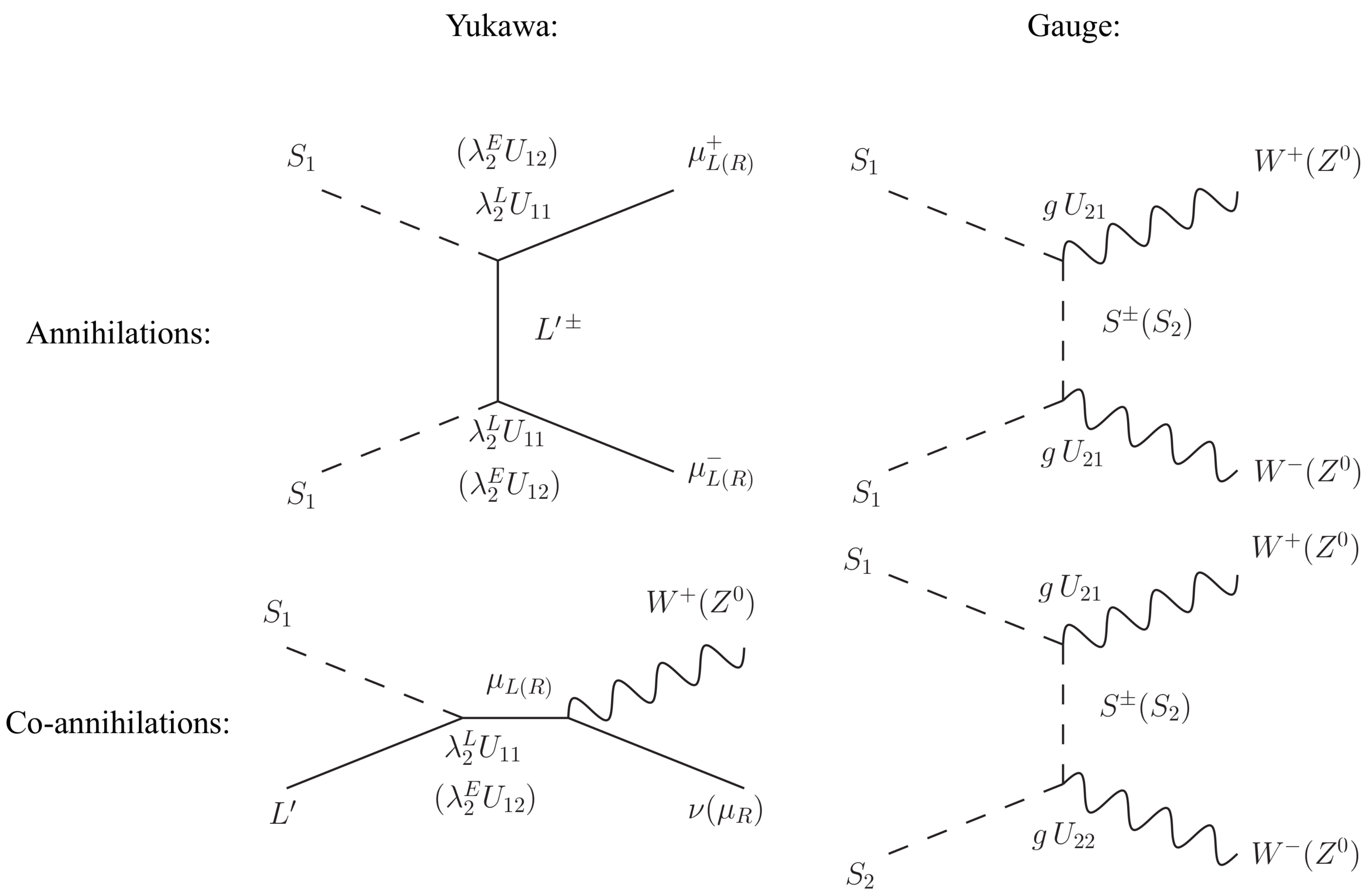}
\caption{(Co-)annihilation processes relevant for $S_1$ DM.\label{fig:DMdiag}}
\end{figure}

As discussed above, the extra fields we introduce are assumed to be odd under an unbroken $\mathbf Z_2$ parity, which ensures that the lightest new state is stable. In the following, we are considering the case that such state is neutral so that it can provide a candidate of dark matter.
In particular, we focus on the lightest scalar $S_1$. 
Furthermore, we assume thermal dark matter production, i.e.~that the standard frieze-out mechanism is at work. 
Despite the reduced field content of the model, a substantial number of annihilation and co-annihilation processes can control the DM relic density. 
Some of the most relevant modes are depicted in \fref{fig:DMdiag}.
The relative importance of a single process depends on the size of the new couplings, as well as on the nature of the DM candidate $S_1$ that, we remind, is a mixture of a SM-singlet scalar and the neutral component of a scalar $SU(2)_L$ doublet: $S_1 = U_{11} S^0_s + U_{21} S^0_d$ (cf.~\sref{sec:model}).
In particular, we can identify the three following regimes with distinctive features.
\begin{enumerate}
\item[(i)] $S_1$ is mainly singlet, hence typically over-produced unless some efficient (co-)annihilation process is at work. The observed relic density can be then obtained through Yukawa-controlled processes like those shown in the first column of \fref{fig:DMdiag}, if the new Yukawa interactions are sizeable and the vectorlike fermions are not too heavy and/or the $S_1$-$L^\prime\,(Q^\prime)$ mass difference is small (for the co-annihilation processes). A different possibility is that DM efficiently annihilates via a resonant s-channel Higgs exchange: this can occur if $M_{S_1} \approx m_h/2$ and $S_1$ has a (albeit small) doublet component, cf.~\eref{eq:Lmix}.
\item[(ii)] $S_1$ is mainly doublet. In this case gauge processes as those of the second column of \fref{fig:DMdiag} are very efficient in depleting the DM density in the early universe. If $S_1$ is a pure doublet, the relic density matches the value observed today, 
$\Omega_{\rm DM}h^2 \simeq 0.12$ \cite{Ade:2015xua}, if $m_{S_1}\approx 540$ GeV (see e.g.~\cite{Cirelli:2005uq}), while a lighter $S_1$ would be a subdominant DM component.
\item[(iii)] If $S_1$ is a substantial mixture of both singlet and doublet components, all kind of processes of \fref{fig:DMdiag} are in principle active, as well as annihilations mediated by the Higgs. This scenario is naturally achieved for a moderate mass splitting of the singlet and doublet scalars: as a consequence all the new scalars are close in mass and the gauge co-annihilation modes are particularly effective.  The observed relic abundance is thus easily obtained if $M_S \approx M_D$.
\end{enumerate}

DM direct detection experiments are sensitive to our scenario. Indeed, a Higgs-mediated DM-nucleon interaction (arising from the $S_1 S_1 h$ coupling on one side, and the Higgs coupling to gluons through a top loop, on the other side) can induce a sizeable spin-independent (SI) cross section. The $S_1 S_1 h$ interaction arises from the mixing of the singlet and the doublet and requires substantial components of both in $S_1$, in order to be effective: as we can see from \eref{eq:Lmix}, the coupling is proportional to $a_H U_{21} U_{11}$. Thus we expect direct detection experiments to best constrain the large mixing case (iii).
Moreover, if $S_1$ is mainly doublet, as in case (ii), or through singlet-doublet mixing in the other cases, it can interact with the $Z$ boson.
Thus a tree-level $Z$ exchange may induce a scattering cross section with nuclei several orders of magnitude larger than the present limits. 
However, notice that  the term $Z_\mu S_1 \partialLR S_1$ in \eref{eq:Lgauge} only couples the 
CP-even to the CP-odd component of $S_1$, thus leading to an inelastic DM-nucleus scattering. A mass splitting of just $\mathcal{O}(100)$ keV
between real and imaginary part of $S_1$ (naturally achieved via the quartic couplings in the scalar potential)
then is sufficient to kinematically forbid $Z$-mediated scattering with nuclei \cite{TuckerSmith:2004jv}. In the following, we are assuming that this is the case and only focus on Higgs-mediated elastic DM-nucleon interactions.

Finally, we comment about another possible DM candidate in our model: the neutral component of $L^\prime$. This would constitute a pure fermion doublet DM candidate, akin to a supersymmetric Higgsino. There are two difficulties related to this possibility. First of all, as in the case of Higgsino DM, the observed relic abundance would require $M_L \approx 1.1$ TeV and all the other particles of course heavier than this. The spectrum would be thus too heavy to account for all the flavour effects we are interested in (in particular $b\to s\mu\mu$), as it will appear clear from the quantitative discussion in the rest of this section. The second problem is that $L^{\prime\,0}$ interacts with the $Z$ boson, cf.~the second line of \eref{eq:Lgauge}. As discussed above, an unacceptably large scattering cross section with nuclei can be avoided if a small Majorana mass term splits the Dirac fermion into two Majorana states, e.g.~through mixing with another Majorana fermion (like in the Higgsino-Bino system) but an extension of the model would be required. For these reasons we are not going to consider this possibility further.

In the following, we will numerically calculate the $S_1$ relic density and SI cross section with nuclei by means of the routine micrOMEGAs \cite{Belanger:2014vza,Belanger:2013oya} and show on our parameter space where $\Omega_{\rm DM}h^2 \simeq 0.12$ \cite{Ade:2015xua} is fulfilled and what are the regions excluded by the latest limit of the XENON1T experiment \cite{Aprile:2018dbl}.

%\paragraph{Combined results.}
\subsection{Combined results}
\label{sec:discussion}
\begin{figure}[t!]
\centering
\includegraphics[height=6.5truecm]{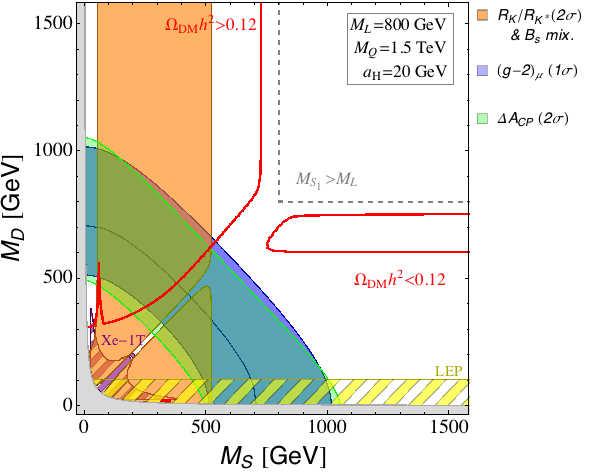}\hfill
\includegraphics[height=6.5truecm]{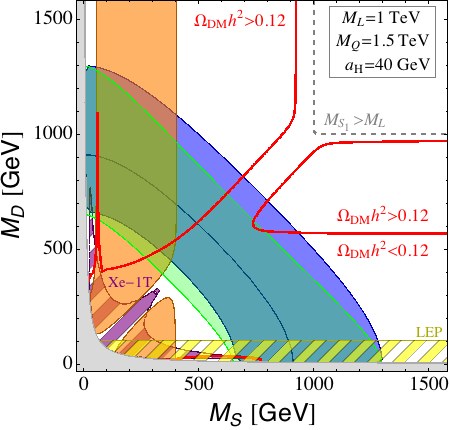} 
\caption{Combined constraints on the singlet-doublet mass plane ($M_S$,\,$M_D$).
The couplings are set to representative values following the discussion in \sref{sec:flavour}:
$\lambda^Q_2 = 0.12$, $\lambda^Q_3 = -2$, $\lambda^L_2 = 1.7$, $\lambda^E_2 = -1$, $\Im(\lambda^U_1) = -1.5$ (left panel), $-1$ (right panel).
All other couplings are set to zero. Cf.~the main text for further details.
  \label{fig:summary1}}
\end{figure}

\begin{figure}[t!]
\centering
\includegraphics[height=6.5truecm]{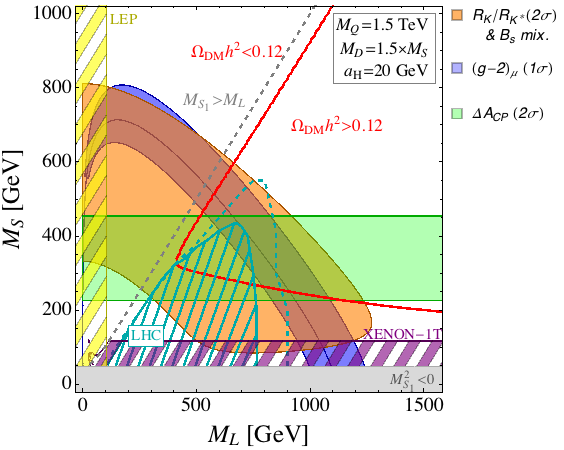}\hfill
\includegraphics[height=6.5truecm]{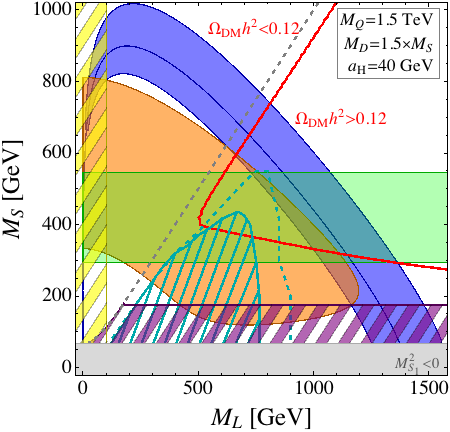}
\caption{Combined constraints on  %the singlet-doublet mass plane ($M_S$,\,$M_D$) (top panels) and
 the vectorlike lepton mass vs the singlet mass ($M_L$,\,$M_S$). The doublet mass is set to $M_D =1.5 \times M_S$.
 The coloured areas and the couplings are as in \fref{fig:summary1}. Cf.~the main text for further details.
  \label{fig:summary2}}
\end{figure}

We end this section discussing the combined impact of the flavour observables presented in \sref{sec:flavour} and the DM/LHC constraints on the parameters of our model. The outcome is summarised in Figures \ref{fig:summary1} and \ref{fig:summary2} for several representative slices of the parameter space. 

In \fref{fig:summary1} we show the singlet-doublet mass plane ($M_S$,\,$M_D$) while setting the mass of the vectorlike quark to a value close to the LHC bound discussed above, $M_Q=1.5$ TeV, and the vectorlike lepton to $M_L=800$ GeV (left panel) and $M_L=1$ TeV (right panel), cf.~the discussion below on the implication of these choices for $\mu^+\mu^-+\met$ searches at the LHC. The couplings are set to values consistent with the findings of \sref{sec:flavour}, as indicated in the caption of \fref{fig:summary1}. 
The coloured areas highlight the portions of the parameter space that are preferred by our flavour observables: in the orange region $b\to s\mu\mu$ data can be fitted within $2\sigma$ simultaneously evading the $B_s$\,--\,$\overline{B}_s$ mixing bound, the green region shows where the observed $\Delta A_{CP}$ is completely accounted for by our NP contribution (at $2\sigma$), while in the blue area the muon $g-2$ discrepancy is solved at the 1$\sigma$ level. The hatched areas are excluded by LEP searches for new charged states  with $M_{S^\pm} \lesssim 100$ GeV \cite{Heister:2002mn,Abdallah:2003xe} (yellow) and the DM direct detection experiment XENON1T (purple).
Besides the value of $M_L$, the main difference between the two panels is the singlet-doublet mixing parameter, set to $a_H= 20$ GeV (left)
and 40 GeV (right). As we can see, by comparing the two plots, a larger value of $a_H$ implies a boost to the effects that depend on the singlet-doublet mixing, such as the chirality-enhanced contributions to the muon $g-2$ and the $\Delta C=1$ chromomagnetic operator, as well as the nucleon-DM interaction.

The line where the $S_1$ relic abundance approximately saturates the observed DM relic density $\Omega_{\rm DM}h^2 = 0.12$ is indicated in red. Given the fact that in both examples the chosen values of the mixing parameter $a_H$ are quite moderate, in the  $M_S < M_D$ region of \fref{fig:summary1}, $S_1$ is typically singlet-dominated and thus in general overabundant: we are in the regime (i) discussed in \sref{sec:DM}. The correct relic density is obtained either due to the Higgs resonance, for $M_S \approx m_h/2$, or when the DM mass approaches the vectorlike lepton mass, in which case the $t$-channel annihilation to muons and the co-annihilations modes become effective (this is the regime where the red line is vertical, i.e.~the relic density has no dependence on the singlet-doublet mixing). 
Notice that, given the LHC bound on $M_Q$, the vectorlike quark does not play a role in DM phenomenology in these examples.
In the figure we can also spot the large mixing regime, labelled as (iii) above: indeed the correct relic density can be achieved on a line close to $M_S \approx M_D$. For $M_S > M_D$ DM mostly behaves as a scalar doublet, i.e.~we are in the regime (ii) of \sref{sec:DM}. 
In this regime, not only the DM annihilation to $W^+W^-$ is very efficient but also, given the moderate values of $M_L$ and the large coupling $\lambda^E_2$, DM annihilation and co-annihilation rates mediated by $L^\prime$ are very large. As a result $S_1$ is typically underabundant. However, the observed relic density $\Omega_{\rm DM}h^2 = 0.12$ can be saturated when the rates of either the gauge or the Yukawa modes decrease to a sufficient extent.

As we can see, in both examples of \fref{fig:summary1}, the red line does overlap with the coloured regions, hence one can find suitable spots where the correct relic density is obtained and all our flavour observables are accounted for.

In \fref{fig:summary2}, we show the effect of varying the vectorlike lepton mass, $M_L$: we plot the same observables as above 
on the ($M_L$,\,$M_S$) plane, while keeping a constant ratio between singlet and doublet masses, $M_D =1.5 \times M_S$. All other parameters are as in \fref{fig:summary1}. In addition, we show the constraint from $\mu^+\mu^-+\met$ searches at the LHC under the simplifying assumption 
BR($L^{\prime\,\pm} \to \mu^\pm S_1$)\,=\,1. The hatched cyan area corresponds to the region excluded by the ATLAS search \cite{Aaboud:2018jiw} as recast in \cite{Calibbi:2018rzv}. The dashed cyan line shows how the bound increases due to the updated analysis in \cite{Aad:2019vnb}: this limit is an estimate based on the excluded production cross section as reported in the auxiliary material of the ATLAS article. 
As we mentioned in \sref{sec:LHC}, we expect that this search can exclude our vectorlike lepton up to $M_L\approx 900$ GeV if 
$M_L \gg M_{S_1}$.\footnote{Still $M_L= 800$ GeV as in the left plot of \fref{fig:summary1} should be viable so long as $M_{S_1}\gtrsim$ 500 GeV.} This is a stringent constraint, but the plots of  \fref{fig:summary2} shows that it does not prevent a simultaneous explanation of
DM and our flavour observables. Indeed, this seems to be possible either for a rather heavy $L^\prime$ (up to $M_L \approx 1.2-1.3$ TeV) or
for the `compressed spectrum' region, where the $L^\prime-S_1$ mass difference is reduced and the LHC searches quickly lose efficiency (because the muons are less energetic) while the correct relic density can be achieved through $L^\prime-S_1$ co-annihilations.
If the latter option may be challenging to test at the LHC (barring perhaps searches for soft leptons as in \cite{Sirunyan:2018iwl}), the former one
will be surely within the sensitivity of future LHC runs.

\section{Conclusions}
\label{sec:conclusions}

We have presented and thoroughly discussed the phenomenological consequences of an extension of the SM, 
featuring a heavy vectorlike quark, a heavy vectorlike lepton, and two scalar fields (a singlet and a doublet) that couple to the Higgs field and hence mix through electroweak symmetry breaking. We have shown that this rather simple setup can provide a simultaneous 
explanation of the $B$-physics anomalies and the muon $g-2$, give a large contribution to the CP violation in charm decays $\Delta A_{CP}$ (to the extent of easily saturating the value recently measured by LHCb), and account for the observed DM abundance, while evading all constraints set by other flavour observables, LHC searches, DM searches.
We found that the novel ingredient of our model (compared to e.g.~\cite{Kawamura:2017ecz,Grinstein:2018fgb}), namely the singlet-doublet mixing, is crucial in order to achieve that. This is because the mixing can give rise to chirally-enhanced dipole transitions that allow to account for the muon $g-2$ and  $\Delta A_{CP}$ for TeV-scale masses of the vectorlike quark and lepton. 
In \sref{sec:couplings}, we have shown the pattern of the new fields' couplings that can address our flavour observables and be compatible with the bounds from other flavour processes. In the spirit of simplified models, we have not discussed how plausible such flavour structure is. It is however encouraging that, at least in the quark sector, the couplings are compatible with a SM-like hierarchical pattern.
Our model could be regarded as a building block of a more complete theory addressing other shortcomings of the Standard Model.Still, it is remarkable that, despite its simplicity, it can consistently account for so many phenomena.
The model can be tested (at least in part) via flavour observables, in particular by the upcoming results of the Fermilab Muon g-2 experiment, and 
future determinations of the SM prediction of the  $B_s$\,--\,$\overline{B}_s$ mass splitting with increased accuracy, and also at the future runs of the LHC, given the necessary presence of charged scalars below 1 TeV and vectorlike fermions in the 1-2 TeV range. 
 Furthermore, a combined explanation of all the phenomena we discussed seems to require the presence of some large Yukawa coupling implying the need of a UV completion below the Landau-pole scale that can be as low as $\approx 100$ TeV.

%%%%%%%%%%%%%%%%%%%%%%%%
\vspace{0.5cm} \noindent {\bf Acknowledgements.}
 We are grateful to Marco Fedele and Federico Mescia for very useful discussions. The Feynman diagrams has been drawn using
 JaxoDraw \cite{Binosi:2003yf,Binosi:2008ig}.
This research was supported by the National Natural Science Foundation of China  under the grants 
No.~11847612 and 11875062 (for TL), 11575151 and 11975195 (for YL), 11805161 (for BZ), by the Natural Science Foundation of Shandong Province under the grants No.~ZR2019JQ004 (YL) and ZR2018QA007 (BZ), and by the Key Research Program 
of Frontier Science, CAS (TL). BZ is also supported by the Basic Science Research Program through the National Research Foundation of Korea (NRF) funded by the Ministry of Education, Science and Technology (NRF-2019R1A2C2003738), and by the Korea Research Fellowship Program through the NRF funded by the Ministry of Science and ICT (2019H1D3A1A01070937).
%%%%%%%%%%%%%%%%%%%%%%%%%%%%%%%%%%%%%%%%%

%%%%%%%%%%%%%%%%%%%%%%%%%%
\newpage

\appendix
\numberwithin{equation}{section}
\section{Lagrangian}
\label{app:lag}

After electro-weak symmetry breaking, the Lagrangian in terms of the mass eigenstates can be written as
%
% \begin{align}
%&{\cal L}  \supset {\cal L}_{\rm mass} + {\cal L}_{\rm mix}   +{\cal L}_{\rm gauge}+  {\cal L}_{\rm yuk}  \, , 
%\end{align}
%with
  ${\cal L} \supset {\cal L}_{\rm mass} + {\cal L}_{\rm mix}   +{\cal L}_{\rm gauge}+  {\cal L}_{\rm yuk}$, where
\begin{align}
&{\cal L}_{\rm mass}  =  - M_Q \overline{U^\prime} U^\prime- M_Q \overline{D^\prime} D^\prime
- M_L \overline{L^{\prime\, 0}} L^{\prime\, 0} - M_L \overline{L^{\prime -}} L^{\prime -} 
  -  M_{S_\alpha}^2 S_\alpha^* S_\alpha  - M_D^2 S^+ S^- \, , 
\\
&{\cal L}_{\rm mix}    =  - \frac{a_H}{\sqrt{2}}\, h  \, U^*_{2\alpha} U_{1 \beta} S^*_\alpha S_\beta  + {\rm h.c.} \,, \label{eq:Lmix}
\\
&   {\cal L}_{\rm gauge}  = 
% photon-fermions, W-fermions
e \,A_\mu \left(       \frac{2}{3} \overline{U^{\prime}} \gamma^\mu U^{\prime } - \frac{1}{3} \overline{D^{\prime}} \gamma^\mu D^{\prime}  
 - \overline{L^{\prime -}} \gamma^\mu L^{\prime -} \right) + \frac{g}{\sqrt2} \left( W_\mu^+ \,\overline{L^{\prime \,0}} \gamma^\mu L^{\prime -} + W_\mu^+ \,\overline{U^{\prime}} \gamma^\mu D^{\prime} + {\rm h.c.}  \right)  \nonumber\\
% Z-fermions
&+ \frac{g}{c_W} Z_\mu \left[ \Big(\frac{1}{2} -\frac{2}{3} s_W^2  \Big) \overline{U^{\prime}} \gamma^\mu U^{\prime}+  \Big(-\frac{1}{2} +\frac{1}{3} s_W^2  \Big) \overline{D^{\prime}} \gamma^\mu D^{\prime}+
 \Big( - \frac{1}{2} + s_W^2  \Big) \overline{L^{\prime -}} \gamma^\mu L^{\prime -} +\frac{1}{2}  \overline{L^{\prime \,0}} \gamma^\mu  L^{\prime \,0} \right]  \nonumber\\
% photon-scalars, Z-scalars
& + i e A_\mu \big(S^+ \partialLR S^-\big) +    \frac{i g}{c_W} Z_\mu \left[  \big( - \tfrac{1}{2} + s_W^2 \big) \big(S^+ \partialLR S^-\big)  + \tfrac{1}{2} U^*_{2 \alpha}U_{2 \beta} \big( S^*_\alpha \partialLR S_\beta\big) \right]
\nonumber     \\
%  W-scalars
& +\frac{ig}{\sqrt{2}} \left[W_\mu^+\,
U^*_{2\alpha} \big(S^*_\alpha  \partialLR S^-\big) +W_\mu^+ S^-  U_{2 \alpha} S_\alpha  \Big( e A^\mu  + \frac{g }{c_W} s_W^2 Z^\mu\Big) + {\rm h.c.}  \right]
  \nonumber     \\
 &  + S^+ S^-   \left[ \tfrac{1}{2} g^2 W^-_\mu W^+_\mu + \Big( e A_\mu  + \frac{g}{c_W} \Big( -\tfrac{1}{2} + s_W^2 \Big) Z_\mu \Big)^2 \right]
 \nonumber\\
& +   \frac{g^2}{8 c_W^2 } U^*_{2 \alpha} U_{2 \beta}\, S^*_\alpha  S_\beta \left( 2 c_W^2 W_\mu^- W_\mu^+ + Z_\mu Z^\mu \right),  \label{eq:Lgauge} \\
%\end{align}
%
%\begin{align}
&{\cal L}_{\rm yuk}   = 
\lambda^L_i  \left( \overline{L^{\prime\, 0}} P_L \nu_i  +  \overline{L^{\prime\, -}} P_L \ell_i  \right)  U_{1\alpha} S_\alpha
+ \lambda^E_i  \left( \overline{L^{\prime\, 0}} P_R \ell_i \,S^+ -     \overline{L^{\prime\, -}} P_R \ell_i \,
 U_{2\alpha}  S_\alpha \right)  \nonumber\\
& + \lambda^Q_i  \left(\overline{U^{\prime}} P_L  u_{j}  +  V^*_{ij} \,  \overline{D^{\prime}} P_L d_j  \right)  U_{1\alpha} S_\alpha
+ \lambda^D_i  \left( \overline{U^{\prime}} P_R d_i \,S^+ - \overline{D^{\prime}} P_R d_i \,
U_{2\alpha}  S_\alpha\right) \nonumber\\
& + \lambda^U_i  \left( \overline{U^{\prime}} P_R u_i\,  U_{2\alpha}  S_\alpha  
+ \overline{D^{\prime}} P_R u_i \,S^- \right) +  {\rm h.c.} \, ,  \end{align}
where $u_i\equiv (u,~c,~t)$, $d_i\equiv (d,~s,~b)$, $\ell_i\equiv (e,~\mu,~\tau)$, and for the LH quarks we chose 
the basis $Q^T_i = (u_{L\,i},~V^*_{ij} \,d_{L\,j})$, see \eref{eq:lambdaQ}, with $V_{ij}$ being elements of the CKM matrix. 
As customary, we defined $c_W\equiv \cos\theta_W$, $s_W\equiv \sin\theta_W$, and 
$\phi_x \partialLR \phi_y \equiv \phi_x (\partial^\mu \phi_y) - (\partial^\mu \phi_x) \phi_y$. 
The expressions for the masses of the neutral scalar eigenstates $S_\alpha$ ($\alpha=1,2$) and the mixing matrix $U$ are given in \sref{sec:model}.
For simplicity, we did not display the terms in the scalar potential (whose coefficients can be assumed to be small enough to have vanishing phenomenological impact apart from providing a mass splitting between CP-even and CP-odd components of $S_\alpha$), apart from the Higgs-scalar coupling in ${\cal L}_{\rm mix}$ arising from the scalar singlet-doublet mixing term.

\section{Wilson coefficients and further observables}

\subsection{$b\to s\ell\ell$}
\label{app:bsll}
We use the following definition of the effective dimension-6 Hamiltonian controlling $b\to s\ell \ell$ transitions (cf.~\cite{Aebischer:2019mlg} and references therein):
\begin{align}
\mathcal{H}^{bs\ell\ell}_{\rm eff} = - \mathcal{N}\,  \sum_x C_{x}^{bs\ell\ell} \mathcal{O}_{x}^{bs\ell\ell}
+{\rm h.c.},
\end{align}
where the normalisation is 
\begin{align}
\label{eq:norm}
\mathcal{N} \equiv \frac{4 G_F}{\sqrt{2}} \frac{e^2}{16\pi^2}V_{tb} V_{ts}^*\,,
\end{align}
and $x$ run over the semi-leptonic operators defined as
\begin{align}	
\mathcal{O}_{9}^{bs\ell\ell} = (\overline{s}\gamma_\mu P_L b) (\overline{\ell} \gamma^\mu \ell),\quad\quad
&\mathcal{O}_{10}^{bs\ell\ell} = (\overline{s}\gamma_\mu P_L b) (\overline{\ell} \gamma^\mu \gamma_5\ell), \nonumber \\
\mathcal{O}_{S}^{bs\ell\ell} = (\overline{s} P_R b) (\overline{\ell} \ell),\quad\quad
&\mathcal{O}_{P}^{bs\ell\ell} = (\overline{s}\gamma_\mu P_R b) (\overline{\ell} \gamma_5 \ell),
\end{align}
and over the $\widetilde{\mathcal{O}}_{x}^{bs\ell\ell}$ operators obtained by exchanging $P_L \leftrightarrow P_R$ in the above expressions.

Within our model, the Wilson coefficients, as obtained from the general results presented in~\cite{Arnan:2019uhr}, read:
\begin{align}
\Delta C_9^{bs\mu\mu}&=-\frac{\lambda_{3}^{Q_d} \lambda_{2}^{Q_d\, *}}{128 \pi^2 \mathcal{N}} \sum_{\alpha=1,2} \frac{\left|U_{1 \alpha}\right|^{4}\left|\lambda_{2}^{L}\right|^{2}+\left|U_{1 \alpha}\right|^{2} \left|U_{2 \alpha}\right|^{2}\left|\lambda_{2}^{E}\right|^{2}}{M_{S_{\alpha}}^{2}} F_2\left(\frac{M_{Q}^{2}}{M_{S_{\alpha}}^{2}}, \frac{M_{L}^{2}}{M_{S_{\alpha}}^{2}}\right),
\\
\Delta C_{10}^{bs\mu\mu}&=\frac{\lambda_{3}^{Q_d} \lambda_{2}^{Q_d *}}{128 \pi^2 \mathcal{N}} \sum_{\alpha=1,2} \frac{\left|U_{1 \alpha}\right|^{4}\left|\lambda_{2}^{L}\right|^{2}-\left|U_{1 \alpha}\right|^{2} \left|U_{2 \alpha}\right|^{2}\left|\lambda_{2}^{E}\right|^{2}}{M_{S_{\alpha}}^{2}} F_2\left(\frac{M_{Q}^{2}}{M_{S_{\alpha}}^{2}}, \frac{M_{L}^{2}}{M_{S_{\alpha}}^{2}}\right), \\
\Delta \widetilde{C}_9^{bs\mu\mu}&=-\frac{\lambda_{3}^{D} \lambda_{2}^{D *}}{128 \pi^2 \mathcal{N}} \sum_{\alpha=1,2} \frac{\left|U_{2 \alpha}\right|^{4}\left|\lambda_{2}^{E}\right|^{2}+\left|U_{1 \alpha}\right|^{2} \left|U_{2 \alpha}\right|^{2}\left|\lambda_{2}^{L}\right|^{2}}{M_{S_{\alpha}}^{2}} F_2\left(\frac{M_{Q}^{2}}{M_{S_{\alpha}}^{2}}, \frac{M_{L}^{2}}{M_{S_{\alpha}}^{2}}\right),
\\
\Delta \widetilde{C}_{10}^{bs\mu\mu}&=-\frac{\lambda_{3}^{D} \lambda_{2}^{D *}}{128 \pi^2 \mathcal{N}} \sum_{\alpha=1,2} \frac{\left|U_{2 \alpha}\right|^{4}\left|\lambda_{2}^{E}\right|^{2}-\left|U_{1 \alpha}\right|^{2} \left|U_{2 \alpha}\right|^{2}\left|\lambda_{2}^{L}\right|^{2}}{M_{S_{\alpha}}^{2}} F_2\left(\frac{M_{Q}^{2}}{M_{S_{\alpha}}^{2}}, \frac{M_{L}^{2}}{M_{S_{\alpha}}^{2}}\right),\\
\Delta C_S^{bs\mu\mu}&=-\frac{\lambda_{3}^{D} \lambda_{2}^{Q_d *}}{64 \pi^2 \mathcal{N}} \sum_{\alpha=1,2} \frac{\left|U_{1 \alpha}\right|^{2}\left|U_{2 \alpha}\right|^{2}(\lambda_{2}^{L\,*}\lambda_{2}^{E}+\lambda_{2}^{L}\lambda_{2}^{E\,*})}{M_{S_{\alpha}}^{2}} \frac{M_Q M_L}{M_{S_\alpha}^2}G_2\left(\frac{M_{Q}^{2}}{M_{S_{\alpha}}^{2}}, \frac{M_{L}^{2}}{M_{S_{\alpha}}^{2}}\right),\\
\Delta C_P^{bs\mu\mu}&=\frac{\lambda_{3}^{D} \lambda_{2}^{Q_d *}}{64 \pi^2 \mathcal{N}} \sum_{\alpha=1,2} \frac{\left|U_{1 \alpha}\right|^{2}\left|U_{2 \alpha}\right|^{2}(\lambda_{2}^{L}\lambda_{2}^{E\,*}-\lambda_{2}^{L\,*}\lambda_{2}^{E})}{M_{S_{\alpha}}^{2}} \frac{M_Q M_L}{M_{S_\alpha}^2}G_2\left(\frac{M_{Q}^{2}}{M_{S_{\alpha}}^{2}}, \frac{M_{L}^{2}}{M_{S_{\alpha}}^{2}}\right),\\
\Delta \widetilde{C}_S^{bs\mu\mu}&=-\frac{\lambda_{3}^{Q_d} \lambda_{2}^{D *}}{64 \pi^2 \mathcal{N}} \sum_{\alpha=1,2} \frac{\left|U_{1 \alpha}\right|^{2}\left|U_{2 \alpha}\right|^{2}(\lambda_{2}^{E\,*}\lambda_{2}^{L}+\lambda_{2}^{E}\lambda_{2}^{L\,*})}{M_{S_{\alpha}}^{2}} \frac{M_Q M_L}{M_{S_\alpha}^2}G_2\left(\frac{M_{Q}^{2}}{M_{S_{\alpha}}^{2}}, \frac{M_{L}^{2}}{M_{S_{\alpha}}^{2}}\right),\\
\Delta \widetilde{C}_P^{bs\mu\mu}&=-\frac{\lambda_{3}^{Q_d} \lambda_{2}^{D*}}{64 \pi^2 \mathcal{N}} \sum_{\alpha=1,2} \frac{\left|U_{1 \alpha}\right|^{2}\left|U_{2 \alpha}\right|^{2}(\lambda_{2}^{E}\lambda_{2}^{L\,*}-\lambda_{2}^{E\,*}\lambda_{2}^{L})}{M_{S_{\alpha}}^{2}} \frac{M_Q M_L}{M_{S_\alpha}^2}G_2\left(\frac{M_{Q}^{2}}{M_{S_{\alpha}}^{2}}, \frac{M_{L}^{2}}{M_{S_{\alpha}}^{2}}\right),
\end{align}
where the loop functions are defined as
\begin{align}	
&F_2(x,y)\equiv \frac{1}{(x-1)(y-1)} + \frac{x^2 \log x}{(x-1)^2(x-y)}+ \frac{y^2 \log y}{(y-1)^2(y-x)}\,, \\
&G_2(x,y)\equiv \frac{2}{(x-1)(y-1)} + \frac{2x \log x}{(x-1)^2(x-y)}+ \frac{2y \log y}{(y-1)^2(y-x)}\,.
\end{align}

Besides $b\to s\ell\ell$ transitions, the above operators also contribute to $B_s\to\ell\ell$ decays, such as $B_s\to\mu\mu$. In particular, (pseudo) scalar
operators provide an helicity-enhanced contribution compared to the SM one, controlled by $\mathcal{O}_{10}^{bs\ell\ell}$. From the measured value of $B_s\to\mu\mu$, that agrees with the SM prediction within 1$\sigma$ \cite{Altmannshofer:2017wqy}, one thus obtains the following bound on the scalar coefficients (calculated at the matching scale of 1 TeV)~\cite{Arnan:2019uhr}:
\begin{align}
\label{eq:Bs}
|\Delta {C}_{S,P}^{bs\mu\mu}|,\,|\Delta \widetilde{C}_{S,P}^{bs\mu\mu}| \lesssim 0.03 \quad(2\sigma).
\end{align}
If scalar or right-handed current operators are not substantially deflected with respect to the SM (as in the case we focus on), the only non-standard contribution to $B_s\to\mu\mu$ is given by $\Delta C_{10}^{bs\mu\mu}$:
\begin{align}
\frac{{\rm BR} (B_s\to \mu^+\mu^-)}{{\rm BR} (B_s\to \mu^+\mu^-)^\text{SM}} = \left| 1+ \frac{\Delta C_{10}^{bs\mu\mu}}{C_{10,\,\text{SM}}^{bs\mu\mu}}  \right|^2 = 0.84 \pm 0.12,
\end{align}
where the allowed range is obtained from \cite{Beneke:2017vpq,Tanabashi:2018oca}. This translates to the 2$\sigma$ constraint
$-0.16 \lesssim \Delta C_{10}^{bs\mu\mu} \lesssim 0.92$, 
which is always fulfilled for the set of parameters satisfying the global fits to $b\to s\ell\ell$ data, cf.~\eref{eq:C9minusC10} and the discussion below it.

\subsection{$b\to s\gamma$}
\label{app:bsg}

This kind of transitions can be accounted for by adding the following electro- and chromo-magnetic dipole operators to the above Lagrangian:
\begin{align}	
\mathcal{O}_{7}^{bs} =\frac{m_b}{e} \, \overline{s} \sigma_{\mu\nu} F^{\mu\nu} P_R b,\quad\quad
&\mathcal{O}_{8}^{bs} = \frac{m_b}{e^2}g_s \, \overline{s} \sigma_{\mu\nu} T^a G_a^{\mu\nu} P_R b,
\end{align}
plus  the corresponding $\widetilde{\mathcal{O}}_{x}^{bs}$ operators obtained by exchanging $P_L \leftrightarrow P_R$.

The leading (chirally-enhanced) contributions of our new fields to the coefficients of the dipole operators are \cite{Arnan:2019uhr}:
\begin{align}
\Delta C^{bs}_7 &\simeq - \frac{2 M_Q}{3\mathcal{N} m_b} \lambda^{Q_d}_{3} \lambda^{D*}_{2}\sum_{\alpha=1,2}   \frac{U^*_{2\alpha} U_{1\alpha}  }{M_{S_\alpha}^2}\, G_8\left(\frac{M^2_Q}{M^2_{S_\alpha}} \right), \\
\Delta \widetilde{C}^{bs}_7 &\simeq - \frac{2 M_Q}{3\mathcal{N} m_b} \lambda^D_{2} \lambda^{Q_d\,*}_{3}\sum_{\alpha=1,2}   \frac{U^*_{1\alpha} U_{2\alpha}  }{M_{S_\alpha}^2}\, G_8\left(\frac{M^2_Q}{M^2_{S_\alpha}} \right), \\
\Delta C^{bs}_8 &\simeq  \frac{2 M_Q}{\mathcal{N} m_b} \lambda^{Q_d}_{3} \lambda^{D*}_{2}\sum_{\alpha=1,2}   \frac{U^*_{2\alpha} U_{1\alpha}  }{M_{S_\alpha}^2}\, G_8\left(\frac{M^2_Q}{M^2_{S_\alpha}} \right), \\
\Delta \widetilde{C}^{bs}_8 &\simeq  \frac{2 M_Q}{\mathcal{N} m_b} \lambda^D_{2} \lambda^{Q_d\,*}_{3}\sum_{\alpha=1,2}   \frac{U^*_{1\alpha} U_{2\alpha}  }{M_{S_\alpha}^2}\, G_8\left(\frac{M^2_Q}{M^2_{S_\alpha}} \right), 
\end{align}
where
\begin{align}
G_7(x)  \equiv \frac{x^2 -4x +3+2 x\log x}{8 (x-1)^3} \, ,\quad G_8(x)  \equiv \frac{x^2 -2 x\log x-1}{8 (x-1)^3} \, .
\end{align}
Given the inclusive measurement of $b\to s\gamma$ decays and the SM prediction, the resulting bound (at 1 TeV) is \cite{Arnan:2019uhr}:
\begin{align}
\label{eq:bsg}
|\Delta {C}_{7}^{bs}+0.19 \Delta {C}_{8}^{bs}|,~ \lesssim 0.06 \quad(2\sigma),
\end{align}
while the constraints on $\Delta \widetilde{C}^{bs}_{7,8}$ are somewhat weaker, due to no interference with the SM contribution.

\subsection{$b\to s\nu\nu$}
\label{app:bsnunu}

Following \cite{Buras:2014fpa}, we employ the effective Hamiltonian
\begin{align}
\mathcal{H}^{bs\nu\nu}_{\rm eff} = - \mathcal{N}\, \left[ 
C^{bs\nu\nu}_L(\overline{s}\gamma_\mu P_L b) (\overline{\nu} \gamma^\mu (1-\gamma_5) \nu) +
C^{bs\nu\nu}_R(\overline{s}\gamma_\mu P_R b) (\overline{\nu} \gamma^\mu (1-\gamma_5) \nu) 
\right]
+{\rm h.c.}\,,
\end{align}
where $\mathcal{N}$ is as in \eref{eq:norm}.
Our model's fields contribute to the above operators as follows: 
\begin{align}
\Delta C_L^{bs\nu\nu}&=-\frac{\lambda_{3}^{Q_d} \lambda_{2}^{Q_d\, *}}{128 \pi^2 \mathcal{N}} \sum_{\alpha=1,2} \frac{\left|U_{1 \alpha}\right|^{4}\left|\lambda_{2}^{L}\right|^{2}}{M_{S_{\alpha}}^{2}} F_2\left(\frac{M_{Q}^{2}}{M_{S_{\alpha}}^{2}}, \frac{M_{L}^{2}}{M_{S_{\alpha}}^{2}}\right), \\
\Delta C_R^{bs\nu\nu}&=-\frac{\lambda_{3}^{D} \lambda_{2}^{D *}}{128 \pi^2 \mathcal{N}} \sum_{\alpha=1,2} \frac{\left|U_{1 \alpha}\right|^{2} \left|U_{2 \alpha}\right|^{2}\left|\lambda_{2}^{L}\right|^{2}}{M_{S_{\alpha}}^{2}} F_2\left(\frac{M_{Q}^{2}}{M_{S_{\alpha}}^{2}}, \frac{M_{L}^{2}}{M_{S_{\alpha}}^{2}}\right)\,.
\end{align}
Given the constraints on RH currents form $B_s$\,--\,$\overline{B}_s$ mixing and the fit to $b\to s\ell\ell$ data, we work in the limit of vanishing 
$\lambda_{i}^{D}$ couplings, resulting in $\Delta C_R^{bs\nu\nu}\approx 0$.
In this limit, one simply finds that \cite{Arnan:2019uhr}
\begin{align}
\frac{{\rm BR}(B\to K \nu\nu)}{{\rm BR}(B\to K \nu\nu)^{\rm SM}} \approx \frac{{\rm BR}(B\to K^* \nu\nu)}{{\rm BR}(B\to K^* \nu\nu)^{\rm SM}}
\approx \frac{2\left|C^{bs\nu\nu}_{L,{\rm SM}}\right|^2+\left|C^{bs\nu\nu}_{L,{\rm SM}} + \Delta C_L^{bs\nu\nu}\right|^2}{3\left|\Delta C_L^{bs\nu\nu}\right|^2},
\end{align}
as the measurement can not distinguish among neutrino flavours. The SM prediction for the Wilson coefficient is numerically given by \cite{Buras:2014fpa}
\begin{align}
C^{bs\nu\nu}_{L,{\rm SM}}  \simeq -1.47 / \sin^2\theta_W.
\end{align}

\subsection{Meson mixing}
\label{app:DeltaM}
We work with the following $\Delta B=2$ effective Hamiltonian:
\begin{align}
\mathcal{H}^{bd_i}_{\rm eff} \supset \sum_x C_x^{bd_i} \, \mathcal{O}^{bd_i}_x  +{\rm h.c.}\,,
\quad\quad {\rm with}~~d_i=d,s.
\end{align}
The operators are defined as 
\begin{align}
&\mathcal{O}^{bd_i}_1 =(\overline{d_i^a}\gamma_\mu P_L b^a)( \overline{d_i^b}\gamma^\mu P_L b^b)\, ,
&\widetilde{\mathcal{O}}^{bd_i}_1 =(\overline{d_i^a}\gamma_\mu P_R b^a)( \overline{d_i^b}\gamma^\mu P_R b^b)\, , \\
&\mathcal{O}^{bd_i}_2 =(\overline{d_i^a}  P_L b^a)( \overline{d_i^b} P_L b^b)\,, 
&\widetilde{\mathcal{O}}^{bd_i}_2 =(\overline{d_i^a}  P_R b^a)( \overline{d_i^b} P_R b^b)\,, \\
&\mathcal{O}^{bd_i}_3 =(\overline{d_i^a}  P_L b^b)( \overline{d_i^b} P_L b^a)\,,
&\widetilde{\mathcal{O}}^{bd_i}_3 =(\overline{d_i^a}  P_R b^b)( \overline{d_i^b} P_R b^a)\,, \\
&\mathcal{O}^{bd_i}_4 =(\overline{d_i^a}  P_L b^a)( \overline{d_i^b} P_R b^b)\, , 
&\mathcal{O}^{bd_i}_5 =(\overline{d_i^a}  P_L b^b)( \overline{d_i^b} P_R b^a)\,,
\end{align}
where $a,~b$ are (summed-over) colour indices.

Using the results of~\cite{Arnan:2019uhr} we find for our model's contributions to the above coefficients (at the matching scale $\Lambda\approx 1$ TeV):
 \begin{align}
\Delta  C_1^{bd_i} &= \frac{(\lambda^{Q_d}_3 \lambda^{Q_d\,*}_{i})^2}{128\pi^2} \sum_{\alpha=1,2}\frac{|U_{1\alpha}|^4}{M_{S_\alpha}^2} 
F\left(\frac{M^2_Q}{M_{S_\alpha}^2}\right),
\\
\Delta  C_2^{bd_i} &= \frac{(\lambda^{Q_d}_3 \lambda^{D\,*}_{i})^2}{64\pi^2} \sum_{\alpha=1,2}\frac{|U_{1\alpha}|^2 |U_{2\alpha}|^2}{M_{S_\alpha}^2} 
\frac{M_{Q}^2}{M_{S_\alpha}^2}G\left(\frac{M^2_Q}{M_{S_\alpha}^2}\right),
\\
\Delta  C_4^{bd_i} &= \frac{(\lambda^{Q_d}_3\lambda^D_3 \lambda^{Q_d\,*}_{i}\lambda^{D\,*}_{i})}{32\pi^2} \sum_{\alpha=1,2}\frac{|U_{1\alpha}|^2 |U_{2\alpha}|^2}{M_{S_\alpha}^2} 
\frac{M_{Q}^2}{M_{S_\alpha}^2}G\left(\frac{M^2_Q}{M_{S_\alpha}^2}\right),
\\
\Delta  C_5^{bd_i} &=- \frac{(\lambda^{Q_d}_3\lambda^D_3 \lambda^{Q_d\,*}_{i}\lambda^{D\,*}_{i})}{32\pi^2} \sum_{\alpha=1,2}\frac{|U_{1\alpha}|^2 |U_{2\alpha}|^2}{M_{S_\alpha}^2} 
\frac{M_{Q}^2}{M_{S_\alpha}^2}G\left(\frac{M^2_Q}{M_{S_\alpha}^2}\right),
\\
\Delta  \widetilde{C}_1^{bd_i} &= \frac{(\lambda^D_3 \lambda^{D\,*}_{i})^2}{128\pi^2} \sum_{\alpha=1,2}\frac{|U_{2\alpha}|^4}{M_{S_\alpha}^2} 
F\left(\frac{M^2_Q}{M_{S_\alpha}^2}\right),
\\
\Delta  \widetilde{C}_2^{bd_i} &= \frac{(\lambda^D_3 \lambda^{Q_d\,*}_{i})^2}{64\pi^2} \sum_{\alpha=1,2}\frac{|U_{1\alpha}|^2 |U_{2\alpha}|^2}{M_{S_\alpha}^2} 
\frac{M_{Q}^2}{M_{S_\alpha}^2}G\left(\frac{M^2_Q}{M_{S_\alpha}^2}\right),
\\
\Delta  C_3^{bd_i} &= \Delta  \widetilde{C}_3^{bd_i} = 0\,,
\end{align}
where
\begin{align}
F(x)\equiv \frac{x^2-1-2x\log x}{(x-1)^3},\quad G(x)\equiv \frac{2x-1-(x+1)\log x}{(x-1)^3}.
\end{align}

The $D$\,--\,$\bar{D}$ mixing operators are easily obtained by taking the ones written above for $B$\,--\,$\bar{B}$ mixing with $d,s\rightarrow u$ and $b\rightarrow c$. Our model's contributions to the coefficients are thus:
 \begin{align}
\Delta  C_1^{cu} &= \frac{(\lambda^{Q_u}_2 \lambda^{Q_u\,*}_{1})^2}{128\pi^2} \sum_{\alpha=1,2}\frac{|U_{1\alpha}|^4}{M_{S_\alpha}^2} 
F\left(\frac{M^2_Q}{M_{S_\alpha}^2}\right),
\\
\Delta  C_2^{cu} &= \frac{(\lambda^{Q_u}_2 \lambda^{U\,*}_{1})^2}{64\pi^2} \sum_{\alpha=1,2}\frac{|U_{1\alpha}|^2 |U_{2\alpha}|^2}{M_{S_\alpha}^2} 
\frac{M_{Q}^2}{M_{S_\alpha}^2}G\left(\frac{M^2_Q}{M_{S_\alpha}^2}\right),
\\
\Delta  C_4^{cu} &= \frac{(\lambda^{Q_u}_2\lambda^U_2 \lambda^{Q_u\,*}_{1}\lambda^{U\,*}_{1})}{32\pi^2} \sum_{\alpha=1,2}\frac{|U_{1\alpha}|^2 |U_{2\alpha}|^2}{M_{S_\alpha}^2} 
\frac{M_{Q}^2}{M_{S_\alpha}^2}G\left(\frac{M^2_Q}{M_{S_\alpha}^2}\right),
\\
\Delta  C_5^{cu} &=- \frac{(\lambda^{Q_u}_2\lambda^U_2 \lambda^{Q_u\,*}_{1}\lambda^{U\,*}_{1})}{32\pi^2} \sum_{\alpha=1,2}\frac{|U_{1\alpha}|^2 |U_{2\alpha}|^2}{M_{S_\alpha}^2} 
\frac{M_{Q}^2}{M_{S_\alpha}^2}G\left(\frac{M^2_Q}{M_{S_\alpha}^2}\right),
\\
\Delta  \tilde{C}_1^{cu} &= \frac{(\lambda^U_2 \lambda^{U\,*}_{1})^2}{128\pi^2} \sum_{\alpha=1,2}\frac{|U_{2\alpha}|^4}{M_{S_\alpha}^2} 
F\left(\frac{M^2_Q}{M_{S_\alpha}^2}\right),
\\
\Delta  \tilde{C}_2^{cu} &= \frac{(\lambda^U_2\lambda^{Q_u\,*}_{1})^2}{64\pi^2} \sum_{\alpha=1,2}\frac{|U_{1\alpha}|^2 |U_{2\alpha}|^2}{M_{S_\alpha}^2} 
\frac{M_{Q}^2}{M_{S_\alpha}^2}G\left(\frac{M^2_Q}{M_{S_\alpha}^2}\right).
\end{align}

\subsection{Neutron EDM}
\label{app:dn}
Following \cite{Konig:2014iqa}, we define the relevant electro-magnetic and chromo-magnetic dipole operators as
\begin{align}
-\mathcal{H}_{\rm eff} \supset C^q_\gamma~ \frac{m_q e}{16\pi^2} \, \overline{q} \sigma_{\mu\nu} F^{\mu\nu} P_R q
+C^q_g ~\frac{m_q g_s}{16\pi^2} \, \overline{q} \sigma_{\mu\nu} T^a G_a^{\mu\nu} P_R q +{\rm h.c.}\,,
\quad\quad {\rm with}~~q=u,\,d.
\end{align}
The electric dipole moment (EDM) and the chromo-eletric dipole moment (CEDM) of (up and down) quarks (respectively $d_q$ and $\tilde{d}_q$) are simply given by the imaginary part of the coefficients of above operators, namely:
\begin{align}
d_q =  \frac{m_q e}{8\pi^2}~ \Im (C^q_\gamma(m_n)),\quad\quad \tilde{d}_q =  \frac{m_q g_s}{8\pi^2}~ \Im (C^q_g(m_n)).
\end{align}
In our model the leading (chirally-enhanced) contributions to the Wilson coefficients of the dipole operators read:
\begin{align}
C^u_\gamma(\Lambda) &\simeq  -\frac{4 M_Q}{3 m_u} \lambda^U_{1} \lambda^{Q_u\,*}_{1}\sum_{\alpha=1,2}   \frac{U_{2\alpha} U^*_{1\alpha}  }{M_{S_\alpha}^2}\, G_\gamma\left(\frac{M^2_Q}{M^2_{S_\alpha}} \right), \\
C^u_g(\Lambda)&\simeq  \frac{2M_Q}{m_u} \lambda^U_{1} \lambda^{Q_u\,*}_{1}\sum_{\alpha=1,2}   \frac{U_{2\alpha} U^*_{1\alpha}  }{M_{S_\alpha}^2}\, G_g\left(\frac{M^2_Q}{M^2_{S_\alpha}} \right), \\
C^d_\gamma(\Lambda) &\simeq  \frac{2 M_Q}{3 m_d} \lambda^D_{1} \lambda^{Q_u\,*}_{1}\sum_{\alpha=1,2}   \frac{U_{2\alpha} U^*_{1\alpha}  }{M_{S_\alpha}^2}\, G_\gamma\left(\frac{M^2_Q}{M^2_{S_\alpha}} \right), \\
C^d_g(\Lambda) &\simeq  \frac{2M_Q}{m_d} \lambda^D_{1} \lambda^{Q_u\,*}_{1}\sum_{\alpha=1,2}   \frac{U_{2\alpha} U^*_{1\alpha}  }{M_{S_\alpha}^2}\, G_g\left(\frac{M^2_Q}{M^2_{S_\alpha}} \right).
\end{align}
where
\begin{align}
G_\gamma(x)  \equiv \frac{x^2 -4x +3+2 x\log x}{8 (x-1)^3} \, ,\quad G_g(x)  \equiv \frac{x^2 -2 x\log x-1}{8 (x-1)^3} \, .
\end{align}

The QCD running, calculated as in \cite{Buchalla:1995vs,Konig:2014iqa}, numerically gives: 
$C^{u,d}_\gamma (m_n) =  0.52\times C^{u,d}_\gamma (1\,{\rm TeV})+0.11\times C^{u,d}_g (1\,{\rm TeV}) $,
$C^{u,d}_g (m_n) =  0.52\times C^{u,d}_g (1\,{\rm TeV})$.

According to the QCD sum rule calculation of \cite{Pospelov:2000bw} the resulting neutron EDM is given by:
\begin{align}
\label{eq:nedm}
d_n = (1\pm0.5) \left(1.4\,(d_d -0.25 d_u) +1.1 e\,(\tilde{d}_d +0.5 \tilde{d}_u)  \right).
\end{align}
The experimental limit on this quantity is (at 90\% CL) \cite{Afach:2015sja}:
\begin{align}
|d_n| < 3\times 10^{-26}~e\,{\rm cm}.
\end{align}

As before, we work in the limit of small $\lambda^D_i$ couplings. Therefore, the up-quark contributions $d_u$ and $\tilde{d}_u$ dominate in \eref{eq:nedm} and we can use the experimental limit to set a bound on $\Im(\lambda^U_{1} \lambda^{Q_u\,*}_{1})$, cf.~\sref{sec:DeltaACP}.

\subsection{Electroweak precision observables}
\label{sec:ewpo}

In our model, oblique parameters are affected by the presence of a scalar doublet which does not mix with the Higgs doublet, similarly to the inert doublet
model first proposed in \cite{Barbieri:2006dq}. The only role of the singlet is to induce through mixing a splitting between the mass of the neutral (mainly) doublet state $S_2$ and the charged state $S^{\pm}$ (whose mass is simply $M_D$) that the oblique parameters are sensitive to. Following the discussion in \cite{Barbieri:2006dq} we find
\begin{align}
\Delta S \approx \frac{1}{2\pi}\ln \frac{M_{S_2}^2}{M_D^2}\,,\quad\quad \Delta T \approx \frac{1}{12\pi^2 \alpha v^2}(M_D -M_{S_2})^2\,.
\end{align}
For our parameter space regions of interest, the resulting effects are tiny. For instance, for the benchmark values adopted in \sref{sec:flavour} ($M_S = 350$ GeV, $M_D = 500$ GeV, $a_H = 20$ GeV) we get $\Delta S \approx 10^{-5}$, $\Delta T \approx 10^{-7}$. These results are orders of magnitude smaller than the effects allowed by fits to electroweak data: $\Delta S = 0.00\pm0.07$, $\Delta T = 0.05\pm0.06$ \cite{Tanabashi:2018oca}. This outcome was to be expected, as an $\ord{100}$ GeV mass splitting is required to have an effect of the order $\Delta T\approx 0.1$ \cite{Barbieri:2006dq}, while in our setup the neutral-charged doublet mass splitting is typically tiny, $\approx 0.1$ GeV, for the values of the parameters chosen throughout the paper.

Other precision observables that can be in principle affected by our extra fields are $Z$-pole observables, in particular the $Z$ couplings to fermions.
As explicitly shown in \cite{Arnan:2019uhr}, these are electroweak symmetry breaking effects that are either proportional to $\sim M_Z^2/M^2$, where $M$ is the typical mass scale of the fields running in the loop, or to electroweak mixing in the new physics sector, that is, in our case, the scalar singlet-doublet mixing  via an Higgs vev. This latter contribution is suppressed in our case by a factor $\sim a_H^2 v^2/M^4$, which makes it negligible. We show the explicit result for the deflection of the $Z$ coupling $\Delta g^Z_{\mu_L}$ to LH muons, a field for which an explanation of the $B$-physics anomalies require a large coupling to the scalar singlet and the vectorlike lepton, $\lambda_2^L \gtrsim 1.5$, cf.~\sref{sec:flavour}. The results for the $Z$ coupling to other SM fermions are qualitatively analogous. Following  \cite{Arnan:2019uhr} we find
\begin{align}	
\Delta g^Z_{\mu_L} =& \frac{|\lambda_2^L|^2}{32\pi^2} \left[ \sum_{\alpha,\beta} U_{1\alpha}U_{1\beta}^*\, g^Z_{S_\alpha S_\beta} H_Z\left(\frac{M_L^2}{M_{S_\alpha}^2},\frac{M_L^2}{M_{S_\beta}^2}\right)+ \right. \nonumber \\
& M_Z^2\,\sum_\alpha |U_{1\alpha}|^2 g^Z_{L^\prime} \left( 2\,\frac{M_L^2}{M_{S_\alpha}^4} \widetilde{G}_Z\left(\frac{M_L^2}{M_{S_\alpha}^2}\right)-
\frac{2}{3}\,\frac{1}{M_{S_\alpha}^2} \widetilde{F}_Z\left(\frac{M_L^2}{M_{S_\alpha}^2}\right) \right)  \nonumber\\
& \left. -\frac{1}{3}M_Z^2 \sum_{\alpha,\beta} U_{1\alpha}U_{1\beta}^*\, \frac{g^Z_{S_\alpha S_\beta} }{M_L^2} \widetilde{H}_Z\left(\frac{M_L^2}{M_{S_\alpha}^2},\frac{M_L^2}{M_{S_\beta}^2}\right)
\right]\,,
\label{eq:Zcoupl}
\end{align}  
where the couplings of our extra fields to the $Z$ are $g^Z_{S_\alpha S_\beta} = \frac{g}{2 c_W} U_{2\alpha}U_{2\beta}^*$, $g^Z_{L^\prime} = \frac{g}{2 c_W} \left(-\frac{1}{2} + s_W^2\right)$, and the loop functions read:
\begin{align}
&H_Z\left(x,y\right) =  \frac{y \ln x}{(x-1)(x-y)} + x\leftrightarrow y\,,   \\
&\widetilde{G}_Z\left(x\right) = \frac{2 + 3 x - 6 x^2 + x^3 + 6 x \ln x}{6 x(x-1)^4}\,,\quad 
\widetilde{F}_Z\left(x\right) = \frac{11 - 18 x +9 x^2 -2 x^3 + 6  \ln x}{6 (x-1)^4}\,, \\
&\widetilde{H}_Z\left(x,y\right) =  \left( \frac{x^2y}{(y-1)(x-y)^2} - \frac{x^2y^2(3x-y-2)\ln x}{(x-1)^2(x-y)^3} \right) + x\leftrightarrow y \,.
\end{align}
Notice that the first line of \eref{eq:Zcoupl} (corresponding to terms that are controlled by the singlet-doublet mixing) depends on the
mixing angles as $U_{1\alpha}U_{1\beta}^*U_{2\alpha}U_{2\beta}^*$, which implies the $\sim a_H^2 v^2/M^4$ suppression mentioned above, as one can 
see from the analytical expression for the mixing matrix, \eref{eq:Umix}.
For our typical choices of parameters ($M_S = 350$ GeV, $M_D = 500$ GeV, $M_L = 800$ GeV) these terms are always subdominant unless $a_H\gtrsim 100$
GeV, a value that is much above the amount of mixing we have been considering in the paper.
The dominant terms stem from a $Z$-$L^\prime$-$L^\prime$ vertex and only depend on the singlet components of the scalars, cf.~the second line of \eref{eq:Zcoupl}. They do not require mixing but feature a $\sim M_Z^2/M^2$ suppression.  They can be simplified as
\begin{align}
\Delta g^Z_{\mu_L} \approx  \frac{|\lambda_2^L|^2}{32\pi^2} g^Z_{L^\prime} \frac{M_Z^2}{M_S^2} \widetilde{K}_Z\left(\frac{M_L^2}{M_S^2}\right)\,,\quad
\widetilde{K}_Z(x) = \frac{(x-1) (5 + -22 x + 5 x^2) + 6 (3 x-1) \ln x}{9 (x-1)^4}\,.
\end{align}
For the usual choice of parameters of \sref{sec:flavour}  the result is $\Delta g^Z_{\mu_L} \approx - 10^{-5}$, that is, two orders of magnitude below the sensitivity reached by LEP experiments, which set the constraint $\Delta g^Z_{\mu_L} = (-0.1\pm 1.1)\times 10^{-3}$ \cite{Efrati:2015eaa,Arnan:2019uhr}.

%%%%%%%%%%%%%%%%%%%%%%%%%%%%%%%%%%%%%%%%%%%%%%%%%%%%%%%%%%%%%%%%%%%%%%

\bibliographystyle{JHEP} 
\bibliography{refs}% Produces the bibliography via BibTeX.

\end{document}